\documentclass[a4paper,11pt]{article}
\pdfoutput=1 

\usepackage{jcappub} 

\usepackage[T1]{fontenc} 
\usepackage[utf8]{inputenc}
\usepackage{hyperref}
\usepackage{cprotect}
\usepackage{float}
\usepackage{subcaption}
\usepackage{caption}
\usepackage{times}
\usepackage{layouts}

\title{\boldmath Proca-stinated Cosmology II: Matter, Halo, and Lensing Statistics in the vector Galileon}

\author[a,b]{Christoph Becker}
\author[a]{, Alexander Eggemeier}
\author[a]{, Christopher T. Davies}
\author[a]{ and Baojiu Li}

\affiliation[a]{Institute for Computational Cosmology, Department of Physics, Durham University, South Road, Durham DH1 3LE, United Kingdom}
\affiliation[b]{Institute for Data Science, Durham University, South Road, Durham DH1 3LE, United Kingdom}

\emailAdd{christoph.becker@durham.ac.uk}
\emailAdd{alexander.eggemeier@durham.ac.uk}
\emailAdd{christopher.t.davies@durham.ac.uk}
\emailAdd{baojiu.li@durham.ac.uk}

\abstract{
The generalised Proca (GP) theory is a modified gravity model in which the acceleration of the cosmic expansion rate can be explained by self interactions of a cosmological vector field. In this paper we study a particular sub-class of the GP theory, with up to cubic order Lagrangian, known as the cubic vector Galileon (cvG) model. This model is similar to the cubic scalar Galileon (csG) in many aspects, including a fifth force and the Vainshtein screening mechanism, but with the additional flexibility that the strength of the fifth force depends on an extra parameter -- interpolating between zero and the full strength of the csG model -- while the background expansion history is independent of this parameter. It offers an interesting alternative to $\Lambda$CDM in explaining the cosmic acceleration, as well as a solution to the tension between early- and late-time measurements of the Hubble constant $H_0$. To identify the best ways to test this model, in this paper we conduct a comprehensive study of the phenomenology of this model in the nonlinear regime of large-scale structure formation, using a suite of N-body simulations run with the modified gravity code {\tt ECOSMOG}. By inspecting thirteen statistics of the dark matter field, dark matter haloes and weak lensing maps, we find that the fifth force in this model can have particularly significant effects on the large-scale velocity field and lensing potential at late times, which suggest that redshift-space distortions and weak lensing can place strong constraints on it.
}

\begin{document}
\def\ramses{{\tt RAMSES}}
\def\rayramses{{\tt Ray-Ramses}}
\def\ecosmog{{\tt ECOSMOG}}
\def\rockstar{{\tt ROCKSTAR}}
\def\subfind{{\tt SUBFIND}}
\def\fof{{\tt FoF}}
\def\camb{{\tt CAMB}}
\def\lcdm{{$\Lambda$CDM}}
\def\Mpch{{h^{-1}{\rm Mpc}}}
\def\Msunh{\ {\rm M}_{\odot}/h}
\def\hMpc{\ h/{\rm Mpc}}
\def\deg{\ {\rm deg}}
\def\betathree{\ \tilde{\beta}_3}
\def\nbody{{$N$-body} }
\def\orig{$\kappa$}
\def\ngsn{$N_{\rm GSN}$}
\def\sg{$S_{\rm G}$}
\def\scg{$S_{\rm cG}$}

\maketitle
\flushbottom

\section{Introduction}
\label{sec:intro}

Understanding the laws of physics that govern cosmic structure formation is indispensable for probing into the true nature of gravity, because gravity is the dominant one of the four fundamental forces on cosmological scales. Ever since its establishment, General Relativity (GR) has been a cornerstone of modern cosmology. Even though the predictions of GR have been validated against many tests, these tests are usually limited to small scales such as the solar system \cite{Will:2014kxa}, leaving the cosmological scales underexplored. The current observational results of these latter scales, which trace the dynamics of luminous and dark matter such as stars, galaxies, galaxy clusters, and extended filaments surrounding enormous voids, are generally in good agreement with the current concordance model of cosmology, \lcdm, despite the fact that in recent years a number of tensions between the cosmological parameter estimates from different observational probes have emerged (e.g., \cite{Verde2019,Troxel:2017xyo,Hikage:2018qbn,Asgari2020,Hildebrandt:2018yau}]). However, there is currently no compelling explanation of the smallness of the cosmological constant in this model, which is why alternative models to explain the cosmic acceleration, such as dynamical dark energy and modified gravity (MG), have been widely considered. In particular, in most alternative theories of gravity, the time evolution of large-scale structures can be significantly influenced, so that the observational data in cosmology may allow accurate tests of such models on large scales (for a recent review see \cite{Koyama:2015vza}). 

The last decades have seen many attempts to modify GR. According to the Lovelock theorem, GR is the only theory with second-order local equations of motion for the metric field, which is derivable from a 4-dimensional action \cite{Koyama:2015vza}, and therefore modifications to GR often involve new dynamical degrees of freedom in addition to the metric field, non-locality, higher-dimensional spacetimes and/or higher-order equations. The simplest MG models, for example, usually involve a single scalar degree of freedom with self-interactions or interactions with curvature. It has been well-established that such models can be brought under the umbrella of the Horndeski theory \cite{Horndeski,Kobayashi:2011nu,Deffayet:2011gz}.

One of the most well-known subclasses of the Horndeski theory is the Galileon model \citep{Nicolis:2008in, Deffayet:2009wt, Deffayet:2009mn}, a 4-dimensional effective theory which involves a scalar field with universal coupling to matter and derivative self-interactions. The theory implements Vainshtein screening \cite{Vainshtein} -- a nonlinear mechanism also encountered in theories such as Fierz-Pauli massive gravity \cite{Babichev:2010jd} and the Dvali-Gabadadze-Porrati (DGP) model \cite{Dvali:2000hr} -- to decouple the scalar field from matter near massive objects and therefore can be compatible with Solar system tests of gravity. The model modifies the background expansion history such that it reaches a de Sitter solution in the future without requiring a cosmological constant. Its simplicity makes it possible to study its phenomenology with the help of cosmological $N$-body simulations \cite{Li:2013tda,Barreira:2013eea}. We refer to this model as the scalar Galileon below.

In contrast to the scalar Galileon, the generalised Proca theory (GP) \cite{Heisenberg:2014rta,Allys:2015sht,Jimenez:2016isa}, involves a massive vector field, $A_{\mu}$, with a broken $U(1)$ gauge symmetry and second-order equation of motion (EOM). The theory features Galileon-type derivative self-interactions and couplings to matter. At the background level, the temporal component of the vector field, $A_0$, gives rise to a self-accelerating de Sitter attractor, corresponding to a dark energy equation of state $w_{\rm DE} = -1$ \cite{DeFelice:2016yws}. From the gravitational wave event GW170817 \cite{TheLIGOScientific:2017qsa} with accompanying gamma-ray burst GRB170817A \cite{Goldstein:2017mmi} and other optical counterparts, the speed of propagation of the gravitational waves $c_{T}$ has been tightly constrained to be identical to the speed of light, $c$. This places strong constraints on the allowed operators within the higher order GP Lagrangian. However, even with this restriction, the GP theory is still cosmologicaly interesting, with a theoretically consistent parameter subspace that is free of ghost and Laplacian instabilities \cite{DeFelice:2016yws}, and in which $c_T=c$. 

By introducing non-linear functions into the field Lagrangian of the GP theory to describe its derivative self interactions and couplings with matter, it can be very versatile and flexible. However, in cosmological applications one often specialises to simple choices of these non-linear functions, such as power-law functions, and a number of studies have been conducted along this direction, leading to a good understanding of the cosmological behaviours of the model at background and linear levels. For example, in Refs.~\cite{deFelice:2017paw,Heisenberg:2020xak,Heisenberg:2020xak}, Markov Chain Monte Carlo likelihood analyses were performed for the particular GP theories proposed in Refs.~\cite{DeFelice:2016yws,DeFelice:2016uil}, by exploiting the observational data from type Ia supernovae (SNIa), the cosmic microwave background (CMB), baryonic acoustic oscillations (BAO), the Hubble expansion rate $H(z)$, and redshift-space distortions (RSD). The cross correlation between galaxy field and the integrated Sachs Wolfe (ISW) effect, which has been a powerful probe to constrain the scalar Galileon models, has also been used to constrain parameters of the GP theory \cite{Nakamura:2018oyy}.

In this work, we conduct a broad phenomenological study of a set of five cosmologies based on the toy GP model studied in \cite{Becker:2020azq}. Using the $N$-body code developed in \cite{Becker:2020azq} and augmenting it with an independent set of ray-tracing modules taken from \rayramses \cite{Barreira:2016wqo}, we can supplement previous results with the measurements of non-linear scales and unexplored statistics of the matter field, haloes, and weak lensing. There are several motivations for doing so. One is that we know perturbation theory is not good at quantifying the effects of screening, which is an inherently non-linear phenomenon. $N$-body simulations are the only known tool to accurately study the evolution of the Universe on small, highly non-linear, scales, and can be used to validate or calibrate the predictions of other approaches. Being able to probe small scales will enable us to test a given model against more observational data more accurately, e.g., access scales or regimes that are inaccessible to perturbation theory. For this reason, we will analyse a total of 13 matter, halo and weak lensing statistics, in the effort to identify the ones which are most sensitive to the effect of the fifth force in the GP theory.

This paper is arranged as follows. In Section \ref{sec:proca_theory} we introduce the GP theory and the particular instances of it that we will focus on in this work. In Section \ref{sec:simulations_section} we describe the set up of the $N$-body and ray-tracing simulations on which all following results are based. This is followed by presentations of the main results of the dark matter field (Section \ref{sec:matter_field_statistics}), haloes (Section \ref{sec:halo_statistics}), and weak lensing (Section \ref{sec:lensing_statistics}). Finally, we summarise and discuss in Section \ref{sec:conclusions}.

Throughout this paper, we will use the $(-,+,+,+)$ signature of the metric and abbreviations $\partial A = \partial_{\mu}A^{\mu}$, $(\partial A)^2 = \partial_{\mu}A^{\mu}\partial_{\nu}A^{\nu}$. We set $c=1$ except in expressions where $c$ appears explicitly. Greek indices run over $0,1,2,3$ while Roman indices run over $1,2,3$.

\section{The Generalised Proca (GP) theory}
\label{sec:proca_theory}

In this work, we study the generalised Proca theory of gravity, the most general vector-tensor theories with second-order equations of motion, which contains Lagrangian operators up to cubic order of the Proca field. The action of this model is given by
\begin{equation}\label{eq:proca_general}
    S = \int{\rm d}^4x \sqrt{-g}\left[\mathcal{L}_m + \mathcal{L}_F + \mathcal{L}_2 + \mathcal{L}_3 + \frac{M^2_{\rm Pl}}{2}R\right],
\end{equation}
where $g$ denotes the determinant of the metric tensor $g_{\mu\nu}$, $\mathcal{L}_m$ is the matter Lagrangian density, $\mathcal{L}_{F, 2, 3}$ are the Lagrangian operators introduced by the Proca field, $A_{\mu}$, and the last operator is the standard Einstein-Hilbert term with the Planck mass, $M^{-2}_{\rm Pl} = 8\pi G$, $G$ is Newton's constant, and $R$ is the Ricci scalar. The Proca field can be decomposed as
\begin{equation}
    A_\mu = (A_0,A_i) = (\varphi, B_i + \nabla_i\chi),
\end{equation}
where $\varphi$ is the temporal component of the vector field, $B_i$ is its transverse mode which is divergence free, $\nabla^i B_i = 0$, and $\chi$ is the longitudinal scalar (which can also be referred to as the scalaron field)

The matter Lagrangian density is related to the energy-momentum tensor of a perfect fluid as,
\begin{equation}\label{eq:lagrangian_matter}
    T^{(m)}_{\mu\nu} = -\frac{2}{\sqrt{-g}}\frac{\delta(\sqrt{-g}\mathcal{L}_m)}{\delta g^{\mu\nu}},
\end{equation}
which, assuming that matter is minimally coupled to gravity, satisfies the standard conservation equation
\begin{equation}\label{eq:continuity_eq}
    \nabla^{\mu}T^{(m)}_{\mu\nu} = 0,
\end{equation}
where $\nabla^\mu$ denotes the covariant derivative compatible with $g_{\mu\nu}$.

Introducing the first derivative of the vector field, $B_{\mu\nu}=\nabla_{\mu} A_{\nu}$, we can build the anti-symmetric Faraday tensor as $F_{\mu\nu} \equiv B_{\mu\nu} - B_{\nu\mu}$. The kinetic term of the Proca Lagrangian, $\mathcal{L}_F$, can be described as,
\begin{equation}\label{eq:proca_kinetic_lagrangian}
    \mathcal{L}_F = -\frac{1}{4}F_{\mu\nu}F^{\mu\nu},
\end{equation}
and the self-interaction terms of the vector field are given by,
\begin{equation}\label{eq:lagrangian_cubic}
    \mathcal{L}_2 = G_2(X) = b_{2} X^{p_{2}}, \quad \mathcal{L}_3 = G_3(X)\nabla_{\mu}A^{\mu} = b_{3} X^{p_{3}}\nabla_{\mu}A^{\mu},
\end{equation}
where $X\equiv\frac{1}{2}g_{\mu\nu}A_{\mu}A_{\nu}$, $b_2 \equiv m^2$ is the mass-squared of the vector field that characterises the onset of the acceleration epoch, and $b_{3},p_2,p_3$ are parameters of mass dimension zero in natural units. The choice is generic enough, leaving a viable parameter space in which the theory is free of ghost and Laplacian instabilities \cite{DeFelice:2016yws}. Importantly, due to the derivative self-interaction of the vector field in $\mathcal{L}_3$, the gravitational effect of the field can be screened in dense regions as required by solar system tests. The screening mechanism in this model is analogous to the Vainshtein mechanism \cite{DeFelice:2016cri}. In this work we set $p_2 = p_3 = 1$ as a working example to study the qualitative behaviour of the Proca field and refer to it from now on as cubic vector Galileon (cvG). With this choice, the GP theory behaves as the standard cubic scalar Galileon model (csG) in certain limits \cite{Becker:2020azq}.

When deriving the equation of motions (EOM), we consider the perturbed Friedmann-Robertson-Walker metric in the Newtonian gauge
\begin{equation}\label{eq:FLRW_metric}
    g_{\mu\nu} = -(1+2\Psi){\rm d}t^2 + a^2(t)(1-2\Phi)\delta_{ij}{\rm d}x^i{\rm d}x^j,
\end{equation}
where $a(t)$ is the time-dependent scale factor which is normalised to $a(t_0)=1$ at the present day, and $\delta_{ij}=\text{diag}(+1,+1,+1)$ represents the spatial sector of the background metric that is taken here to be flat, $k=0$.

As shown in \cite{Becker:2020azq}, we expect the `back-reaction' of $B_i$ on the evolution of $\chi$ and $\Phi$ to be very small, justifying the neglect of the $B^i$ field in the simulations. To perform cosmological simulations for this model, we rewrite all required equations in \ecosmog's code units, which we indicate as tilded quantities (details in \cite{Becker:2020azq}). The equations are then rescaled through
\begin{equation}\label{eq:field_perturbation_redefinition}
    \tilde{\chi} = \frac{3\beta_{\rm sDGP}}{2\beta}\tilde{\chi}',
\end{equation}
to make an educated choice of the cvG model parameter possible, by comparing it with the well studied sDGP model (for more details see \cite{Becker:2020azq}). To lighten our notation, we will drop the prime in $\tilde{\chi}'$.

The modified Friedman equation, which depends on the EoM of $\varphi$ at the background level, given by
\begin{equation}\label{eq:phi_cvg}
    \varphi = \frac{H^2_0\tilde{\beta}_2}{3c^2H\tilde{\beta}_3},
\end{equation}
is
\begin{equation}\label{eq:friedmann_code}
    E^2 \equiv \left(\frac{H(a)}{H_0}\right)^2 = \frac{1}{2}\left[\Omega_{m}a^{-3} + \sqrt{\Omega^2_{m}a^{-6} + 4\Omega_P}\right],
\end{equation}
where $H(a)$ is the Hubble expansion rate at $a$, $H_0=H(a=1)$, $\Omega_m$ is the matter density parameter, and $\Omega_{P}$ the Proca field density parameters today,
\begin{equation}\label{eq:omega_cvg}
    \Omega_{P} \equiv 1 - \Omega_{m}.
\end{equation}

We have considered only non-relativistic matter; the inclusion of radiation and massive neutrinos is straightforward. Therefore, the background expansion history in this model is completely determined by $H_0$ and $\Omega_m$.

The modified Poisson equation, rescaled by Eq.~\eqref{eq:field_perturbation_redefinition}, under the quasi-static approximation and in the weak-field limit takes the following form in code units,
\begin{equation}\label{eq:rescaled_poisson_code}
    \tilde{\partial}^2\tilde{\Phi} = \frac{3}{2}\Omega_{m}a\left(\tilde{\rho}-1\right) + \frac{3\beta_{\rm sDGP}}{2\beta}\alpha\tilde{\partial}^2\tilde{\chi}',
\end{equation}
where $\tilde{\rho}$ is the matter density in code unit, $\beta_{\rm sDGP}$ is the coupling strength between matter and the brane-bending mode in the sDGP model, and $\alpha$ and $\beta$ are two time-dependent functions given by
\begin{equation}\label{eq:func_alpha}
    \alpha(a) = \frac{1}{2\sqrt[3]{2}}\betathree^{1/3}\Omega_P^{-1/3}\left[\sqrt{\Omega_m^2a^{-6}+4\Omega_P}-\Omega_ma^{-3}\right],
\end{equation}
and
\begin{equation}\label{eq:func_beta}
    \beta(a) = \frac{1}{2}\left(\frac{\betathree}{2\Omega_P}\right)^{1/3}\left[5\Omega_ma^{-3}+\frac{3\Omega_m^2a^{-6}}{\sqrt{\Omega_m^2a^{-6}+4\Omega_P}}\right] + \betathree,
\end{equation}
which are both fully fixed by specifying $\Omega_m$ and the coupling constant $b_3$ redefined as $\betathree \equiv b_3(8\pi GH_0^2)/c^6$ \cite{Becker:2020azq}.

Finally, the EOM for the longitudinal mode of the Proca field, $\chi$, in the weak-field limit and rescaled by Eq.~\eqref{eq:field_perturbation_redefinition},
\begin{equation}\label{eq:rescaled_chi_eom_code_units}
    \tilde{\partial}^2\tilde{\chi}' + \frac{1}{3\gamma a^4} \left[\left(\tilde{\partial}^2\tilde{\chi}'\right)^2-\left(\tilde{\partial}_i\tilde{\partial}_j\tilde{\chi}'\right)\left(\tilde{\partial}^i\tilde{\partial}^j\tilde{\chi}'\right)\right] = \frac{1}{\beta_{\rm sDGP}}\Omega_{m}a\left(\tilde{\rho}-1\right),
\end{equation}
where the source term on the right-hand side is identical to that in the sDGP equation \cite{Li:2013nua}, and we have defined a new time-dependent function
\begin{equation}
    \gamma(a) \equiv \frac{2\beta^2}{9\beta_{\rm sDGP}R^2_c},
\end{equation}
with the following dimensionless and time-dependent function
\begin{equation}\label{eq:func_Rc2}
    R^2_c(a) = \frac{1}{2}\betathree^{2/3}\left(2\Omega_P\right)^{-2/3}\left[\Omega_ma^{-3}+\sqrt{\Omega_m^2a^{-6}+4\Omega_P}\right].
\end{equation}
Thus, given a matter density field, we can solve for the scalaron field $\chi$ from Eq.~\eqref{eq:rescaled_chi_eom_code_units} and plug it into the modified Poisson equation Eq.~\eqref{eq:rescaled_poisson_code} to solve for $\tilde{\Phi}$. Once $\tilde{\Phi}$ is at hand, we can use finite-difference to calculate the modified gravitational force, which determines how the particles move subsequently. Note that, in this model, $\tilde{\Phi}$ not only determines the geodesics of massive particles, but also those of massless particles such as photons -- in other words, the lensing potential is also modified.

As $\betathree$ is the only ‘free' parameter that enters in all three key equations, it is practical to use it as the model parameter.

\section{Cosmological simulations}
\label{sec:simulations_section}

In this section we present the set of dark-matter-only simulations for five different cosmologies which we use to investigate the phenomenology of the cvG model. Four of these take different values of the model parameter of the cvG model, $\betathree = [10^{-6}, 10^{0}, 10^{1}, 10^{2}]$, and one is their QCDM counterpart\footnote{This is a variant that only considers the modified background expansion history, but uses standard Newtonian gravity to evolve particles}, in the simulation. It is equivalent to the limit $\betathree\rightarrow\infty$ \cite{Becker:2020azq}. To study the cvG effects on the weak lensing (WL) signal, we extended the $N$-body code developed in the previous work \cite{Becker:2020azq} by adding an independent set of ray-tracing modules taken from \rayramses \cite{Barreira:2016wqo}. This allows us to calculate the WL signal ‘on-the-fly' as proposed by \cite{White:1999xa,Li:2010cm}, while taking full advantage of the time and spatial resolution available in the \nbody simulation.

\begin{figure}
    \centering 
    \includegraphics[width=.98\textwidth]{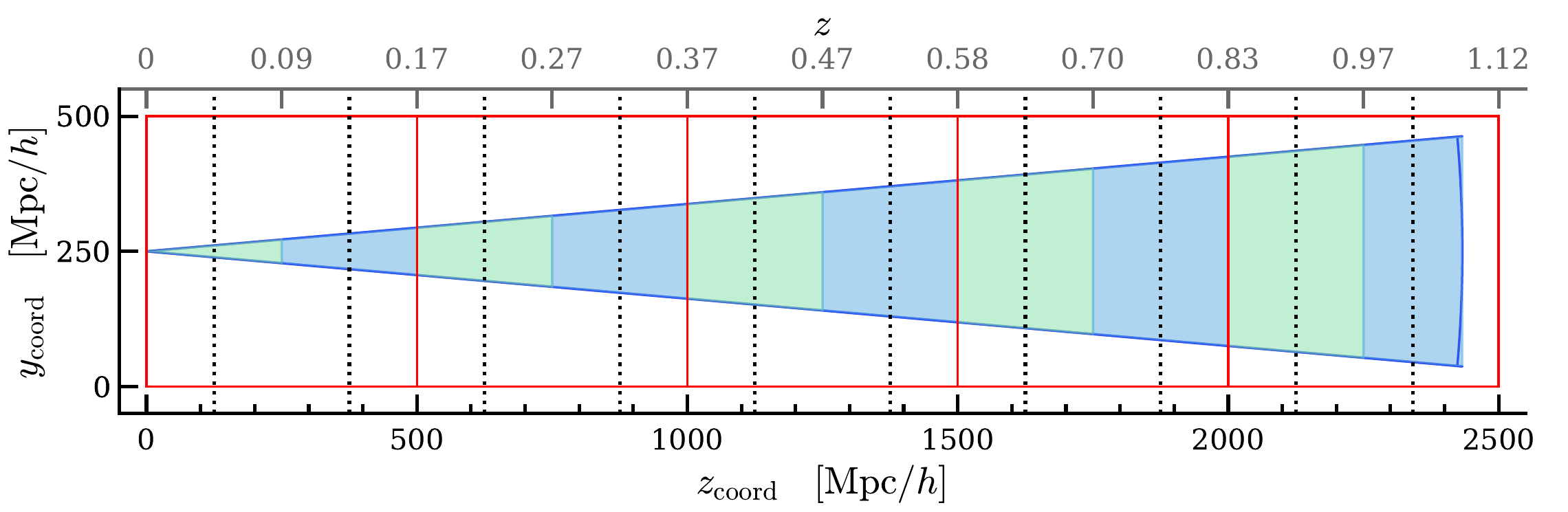}
    \caption{Light-cone layout. The light cone (solid blue line) is made up of five simulation boxes (red squares). All simulated boxes have a side length of $500 \Mpch$ and the light cone has an opening angle of $10 \times 10 \deg^2$. The comoving distance to the observer and redshift are respectively labelled in the lower and upper axes. The vertical dotted lines, which are at distances equal to $1/4$ and $3/4$ times the box size from the nearer side of each box, correspond to the redshifts at which particle snapshots are outputted.}
    \label{fig:lc_setup}
\end{figure}

We construct a light-cone for each cosmology by tiling a set of five simulation boxes, all having an edge-length of $L_{\rm box} = 500 \Mpch$, as shown in Fig.~\ref{fig:lc_setup}. The simulations treats dark matter as collisionless particles described by a phase-space distribution function $f(\mathbf{x}, \mathbf{p}, t)$ that satisfies the Vlasov equation
\begin{equation}\label{eq:vlasov}
    \frac{df}{dt} = \frac{\partial f}{\partial t} + \frac{\mathbf{p}}{m_0 a^2} \cdot \nabla f - m_0 \left(\nabla\Psi\right) \cdot \frac{\partial f}{\partial \mathbf{p}} = 0,
\end{equation}
where $\mathbf{p} = a^2 m_0 \partial\mathbf{x}/\partial t$, $m_0$ is the particle mass, and $\Psi$ is the modified Newtonian potential given by Eq.~\eqref{eq:rescaled_poisson_code}. Note that as we do not include matter species such as photons and neutrinos the two Bardeen potentials are equivalent, $\Psi = \Phi$. Hence to solve $\Psi$, and prior to it the longitudinal Proca mode, via Eq.~\eqref{eq:rescaled_chi_eom_code_units}, they are discretised and evaluated on meshes using the nonlinear Gauss-Seidel relaxation method \cite{Li:2013nua}. The domain grid -- which is the coarsest uniform grid that covers the entire simulation box -- consists of $N_{\rm grid} = 512^3$ cells, which is equal to the number of tracer particles, $N_p$. \ecosmog{} is based on the adaptive-mesh-refinement code \ramses{} \cite{Teyssier:2001cp}, which allows mesh cells in the domain grid to be hierarchically refined -- split into 8 child cells -- when some refinement criterion is satisfied. In our simulations, a cell is refined whenever the effective number of particles inside it exceeds 8. This gives a higher force resolution in dense non-linear regions, where the Vainshtein screening becomes important. The Gauss-Seidel algorithm is run until the difference of the two sides of the PDE, $d^h$, is smaller than a predefined threshold $\epsilon$. We verified that for a value of $\epsilon = 10^{-9} > |d^h|$, the solution of the PDE no longer changes significantly when $\epsilon$ is further reduced.

We use the same set of five different initial conditions (ICs), for each of the five simulations that make up a light-cone for a given cosmology are different, for the different cosmologies. The ICs were generated using {\tt 2LPTic} \cite{Crocce:2006ve}, with cosmological parameters taken from the {Planck Collaboration} \cite{Ade:2015xua},
\begin{equation}\label{eq:cosmological_parameters}
    h = 0.6774, \quad \Omega_{\Lambda} = 0.6911, \quad \Omega_m=0.389, \quad \Omega_B=0.0223, \quad \sigma_8=0.8159.
\end{equation}
The linear matter power spectrum used to generate the ICs is obtained with {\tt CAMB} \cite{2011ascl.soft02026L}. The simulation starts at a relatively low initial redshift $z_{\rm ini}=49$, or $a_{\rm ini}=0.02$, justifying the use of second-order Lagrangian perturbation theory codes such as {\tt 2LPTic}. One possible concern may be that, at this scale factor, differences of matter clustering are already present. However, judging from our experience \cite{Becker:2020azq}, at this time the difference between the growth factors of the cvG model with \lcdm is well below sub-percent level, so that modified effects on the initial matter clustering can be neglected.

\begin{table}
    \caption{Summary of technical details that are identical for all simulations performed for this work. Here $k_{\rm Ny}$ denotes the Nyquist frequency. $\epsilon$ is the residual for the Gauss-Seidel relaxation used in the code \citep{Li:2011vk}, and the two values of the convergence criterion are for the coarsest level and refinements respectively.}
    \begin{center}
        \begin{tabular}{@{}cccccc}
        \hline\hline
         $L_{\rm box}$ & Nr. of particles & $k_{\rm Ny}$ & force resolution & convergence \\
        \hline
         $500$ $\Mpch$ & $512^3$ & $3.21$ $\hMpc$ & $30.52$ $h/{\rm kpc}$ & $|\epsilon|<10^{-12}/10^{-9}$ \\
        \hline
        \end{tabular}
    \label{table:simulations}
    \end{center}
\end{table}

The light-cone, outlined by solid blue lines in Fig.~\ref{fig:lc_setup}, is constructed by positioning the five simulation boxes, outlined by solid red lines in Fig.~\ref{fig:lc_setup}, relative w.r.t. the observer. The geometrical set-up was constructed to place the sources at $z_s = 1$, which is the starting point when the growth rate of matter density perturbations becomes higher than in \lcdm \cite{Becker:2020azq}. The field-of-view (FOV) is set to $10\times10\ {\rm deg}^2$ (so that the wide end of the light-cone is still narrow enough to fit in the simulation box), within which $2048\times2048$ rays are followed by \rayramses{} to compute quantities of interest. \rayramses{} is an on-the-fly ray-tracing code. The rays are initialised when a given simulated box 
reaches a defined redshift (for the closest and furthest box to the observer the initialisation redshift is respectively $z_i = 0.17$ and $z_i = 1.0$), and end after they have traveled the covered length of the box, meaning $500 \Mpch$. As here we are interested in the lensing convergence, $\kappa$, the quantity that is computed along the rays is the two-dimensional Laplacian of the lensing potential, 
\begin{equation}
    \tilde{\nabla}^2 \tilde{\Phi}_{\rm lens, 2{\rm D}} = \tilde{\nabla}_1\tilde{\nabla}^1\tilde{\Phi}_{\rm lens, 2{\rm D}} + \tilde{\nabla}_2\tilde{\nabla}^2\tilde{\Phi}_{\rm lens, 2{\rm D}},
\end{equation}
where 1, 2 denote the two directions on the sky perpendicular to the line of sight (LOS). The values of these two-dimensional derivatives of $\Phi_{\rm lens, 2{\rm D}}$ can be obtained from its values at the centre of the AMR cells via finite differencing and some geometrical considerations (see Refs.~\cite{Li:2010cm, Barreira:2016wqo}). Integrating this quantity as
\begin{equation}\label{eq:kappa}
    \kappa = \frac{1}{c^2}\int_0^{\chi_s}\frac{\chi\left(\chi_s - \chi\right)}{\chi_s} {\tilde{\nabla}}^2\tilde{\Phi}_{\rm lens, 2{\rm D}}(\chi, \vec{\beta}(\chi)) {\rm d}\chi,
\end{equation}
where $c$ is the speed of light, $\chi$ is the comoving distance, $\chi_s$ the comoving distance to the lensing source, and $\vec{\beta}(\chi)$ indicates that the integral is performed along the perturbed path of the photon ($\chi$ is not to be confused with the longitudinal mode of the Proca field). The integral is split into the contribution from each AMR cell that is crossed by a ray, which ensures that the ray integration takes full advantage of the (time and spatial) resolutions attained by the \nbody run. For the WL signal we wish to study in this paper, we employ the Born approximation, in which the lensing signal is accumulated along unperturbed ray trajectories. We will make further notes on the calculations in Sec.~\ref{subsec:weak_lensing}.

\section{Matter field statistics}
\label{sec:matter_field_statistics}

In this section we present the results of various dark matter statistics of the different cvG models and compare them with the predictions by QCDM, to study the impact of the Proca field on these key observables. We start with an analysis of the power spectra in Sect.~\ref{subsec:power_spectra}. In Section \ref{subsec:bispectra}, we consider the leading non-Gaussian statistic in large-scale structure clustering, the bispectrum, which is thus sensitive to deviations from linear evolved perturbations from single field inflation.

To support the analysis and interpretation of the results, we will compare the results of the \nbody simulations to Eulerian standard perturbation theory (SPT), and limit the comparison only to the tree-level statistics. In SPT, the energy and momentum conservation equations can be solved order by order to obtain higher-order corrections to the quantities of interest. The expansion in powers of the linear density field is a simple time dependent scaling of the initial density field (in the Einstein de Sitter approximation),
\begin{equation}\label{eq:spt_delta_scaling}
    \delta(\mathbf{k}, \tau) = \sum^{\infty}_{i=1} D^{n}(\tau)\delta^{(i)}(\mathbf{k}),
\end{equation}
for which the $n$-th order solution is
\begin{equation}\label{eq:spt_delta_order}
    \delta^{(n)}(\mathbf{k}) \sim \int d^3\mathbf{k}_1...d^3\mathbf{k}_n\delta^{({\rm D})}(\mathbf{k} - \mathbf{k}_{1...n})\mathcal{F}_n(\mathbf{k}_1, ..., \mathbf{k}_n)\delta^{(1)}(\mathbf{k}_1, \tau_{\rm ini})...\delta^{(1)}(\mathbf{k}_n, \tau_{\rm ini}),
\end{equation}
with the conformal time $\tau = \int dt/a$, $\mathbf{k}_{1...n} \equiv \mathbf{k}_1 + ... + \mathbf{k}_n$, the density contrast $\delta = \rho/\bar{\rho}$, $\delta^{({\rm D})}$ the 3D Dirac delta function, and $\mathcal{F}_n$ the SPT fundamental mode coupling kernel \cite{Goroff:1986ep,Scoccimarro:1997st}.

When comparing a cvG model to the QCDM counterpart, we do so through their relative difference which we write in short hand as
\begin{equation}\label{eq:relative_difference}
    \frac{\Delta A}{A_{\rm QCDM}} \equiv \frac{A_X - A_{\rm QCDM}}{A_{\rm QCDM}},
\end{equation}
with $A$ a placeholder of the summary statistics, and $X$ will be one of the four cvG models. We calculate $\Delta A / A_{\rm QCDM}$ for each of the five pairs of cvG and QCDM simulations that share the same initial conditions to find its average and $1\sigma$ uncertainty. Taking this ratio removes contributions from cosmic variance, and so its uncertainty is not a direct indicator of how sensitive the various summary statistics are to differences between the cvG models. To provide an estimate of this sensitivity given a survey volume as large as our simulation box, we calculate the signal-to-noise ratio (SNR) of the difference between cvG models and their QCDM counterpart for some summary statistics using the expression
\begin{equation}\label{eq:snr}
    {\rm SNR} \equiv \frac{\Delta A}{\sigma} = \frac{A_X - A_{\rm QCDM}}{\sqrt{\sigma_{X}^2 + \sigma_{\rm QCDM}^2}},
\end{equation}
where $\Delta A$ is the average and $\sigma$ is the standard deviation of the five simulations per cosmological model. However, we note that the SNR values obtained in this way are subject to sample noise, owing to the small number of realisations.

\subsection{Matter and velocity power spectra}
\label{subsec:power_spectra}

To gain insights into the differences of matter clustering and peculiar velocities on linear and nonlinear scales among the various models in this work, we begin our study of dark matter phenomenology by considering the auto power spectra of the matter over-density, $\delta$, given by
\begin{equation}\label{eq:pk_deltadelta}
    \langle \delta(\mathbf{k}_1, t) \delta(\mathbf{k}_2, t) \rangle = (2\pi)^3\delta^{({\rm D})}(\mathbf{k}_1 + \mathbf{k}_2) P_{\delta\delta}(\mathbf{k}_1, t).
\end{equation}
Cosmic structure formation is driven by the spatially fluctuating part of the gravitational potential, $\Phi({\bf x},t)$, in Eq.~\eqref{eq:FLRW_metric}, induced by the density fluctuation $\delta$. In cvG cosmologies we expect an additional boost to the standard gravitational potential with respect to its QCDM counterpart, induced by $\chi$ described by Eq.~\eqref{eq:rescaled_poisson_code}, in regions where the fifth force is not screened by the Vainshtein mechanism. Thus, clustering will be enhanced in the cvG models on some scales, which can be captured by $P_{\delta\delta}$.

\begin{figure}
    \centering 
    \includegraphics[width=.98\textwidth]{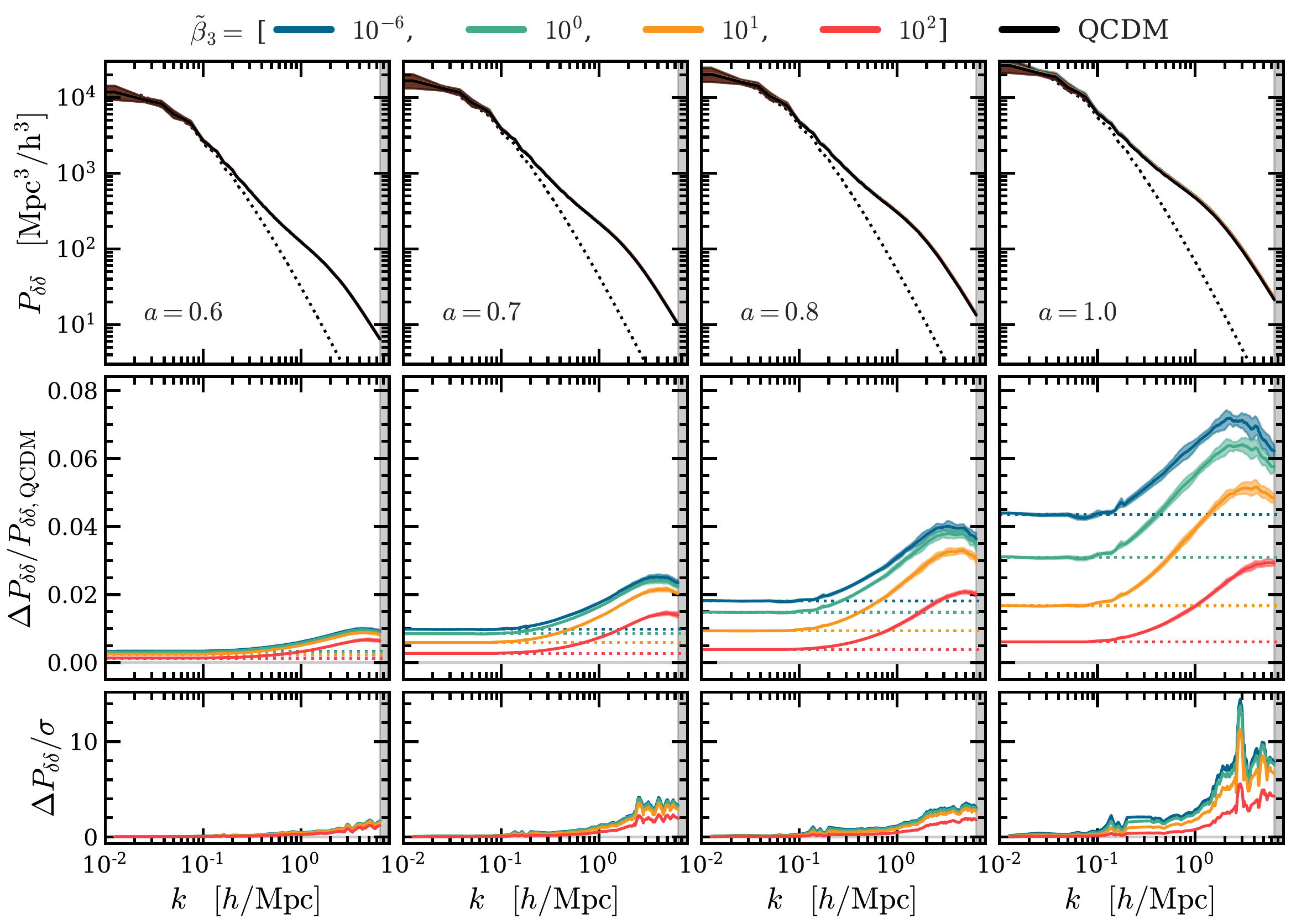}
    \caption{The matter power spectrum in the cvG model. Each column shows the results for a different scale factor: {\it outer left:} $a=0.6$, {\it inner left:} $a=0.7$, {\it inner right:} $a=0.8$, {\it outer right:} $a=1.0$. {\it Top:} The matter power spectrum of linear perturbation theory (dotted) and the cvG model for four values of $\betathree = (10^{-6}, 1, 10, 100)$, indicated by a blue, green, orange and red line respectively. {\it Centre:} Relative differences between the matter power spectra of the cvG and QCDM models. A Savitzky–Golay filter has been used to smooth $\Delta P_{\delta\delta}(k) / P_{\delta\delta,{\rm QCDM}}(k)$ for $k > 0.2 \, \hMpc$. Each panel compares linear perturbation theory (black dotted), to results obtained from full simulation (coloured solid). The vertical grey shaded region in each panel indicates where $k>k_{\rm Ny}$ where $k_{\rm Ny}$ is the Nyquist frequency. {\it Bottom:} The signal-to-noise ratio of the difference between the cvG models and their QCDM counterpart.
    }
    \label{fig:Pk_matter}
\end{figure}

The top row of Fig.~\ref{fig:Pk_matter} compares the linear matter power spectra (black dotted lines) with the simulation results of each cosmology (coloured lines with shaded regions), at $a=0.6$ (outer left), $a=0.7$ (inner left), $a=0.8$ (inner right), and $a=1.0$ (outer right). The linear power spectrum, $P^{(11)}_{\delta\delta}(k;z)$, is obtained by multiplying the initial matter power spectrum at $z_{\rm ini}=49$, $P_{\delta\delta}(k;z_{\rm ini})$, with $\left[D(z)/D(z_{\rm ini})\right]^2$. The nonlinear matter power spectra are measured from particle snapshots using the {\tt POWMES}\footnote{The code is in the public domain, www.vlasix.org/index.php?n=Main.Powmes} code \cite{Colombi:2008dw}. The mean $P_{\delta\delta}$ of the five realisations per cosmology is shown as a coloured line while the standard deviation is indicated as shaded region. The standard deviation is largest at large scales ($k \lesssim 0.1 \hMpc$) due to cosmic variance and the limited simulation size. The vertical shaded region near the right edge of each panel indicates the regime of $k$ beyond the Nyquist frequency\footnote{Note that the Nyquist frequency, $k_{\rm Ny}$, marks the absolute maximum up to which we can the power spectrum can be trusted.}.

The centre row of Fig.~\ref{fig:Pk_matter} shows the relative differences, Eq.~\eqref{eq:relative_difference}, of the matter power spectra. The relative difference has been smoothed to remove noise at scales $k > 0.2 \, \hMpc$, using a Savitzky–Golay filter of second order with a kernel width of $13$ data-points \cite{savitzky1964smoothing}. The power spectrum results agree with the results found in Ref.~\cite{Becker:2020azq} and extend them by including larger scales and measurement uncertainties.

The bottom panel of Fig.~\ref{fig:Pk_matter} shows the SNR of the difference between cvG cosmologies and their QCDM counterpart. From it we can conclude that the SNR is proportional to $k$ while it is inversely related to $\betathree$.

The real-space position of tracers of the matter distribution are not directly measurable, preventing us from comparing $P_{\delta\delta}$ to observations, which rely on the redshift measurement to infer distances. The reason is that peculiar velocities (i.e., additional velocities to the Hubble flow) of the tracers distort the redshift signal along the line of sight. Thus, $P_{\delta\delta}$ is different from its counterpart in redshift space, $P^{\rm s}_{\delta\delta}$, which becomes anisotropic despite the statistical istropy of the Universe; on large scales the two are related by the linear Kaiser formula
\begin{equation}
    P^{\rm s}_{\delta\delta}(k,\mu) = \left(1+f\mu^2\right)^2 P_{\delta\delta}(k),
\end{equation}
where $\mu$ is the angle between the wavevector and the LOS, and $f$ is the linear growth rate defined as $f = {\rm d(ln}\delta) / {\rm d(ln}a)$.

\begin{figure}
    \centering 
    \includegraphics[width=.98\textwidth]{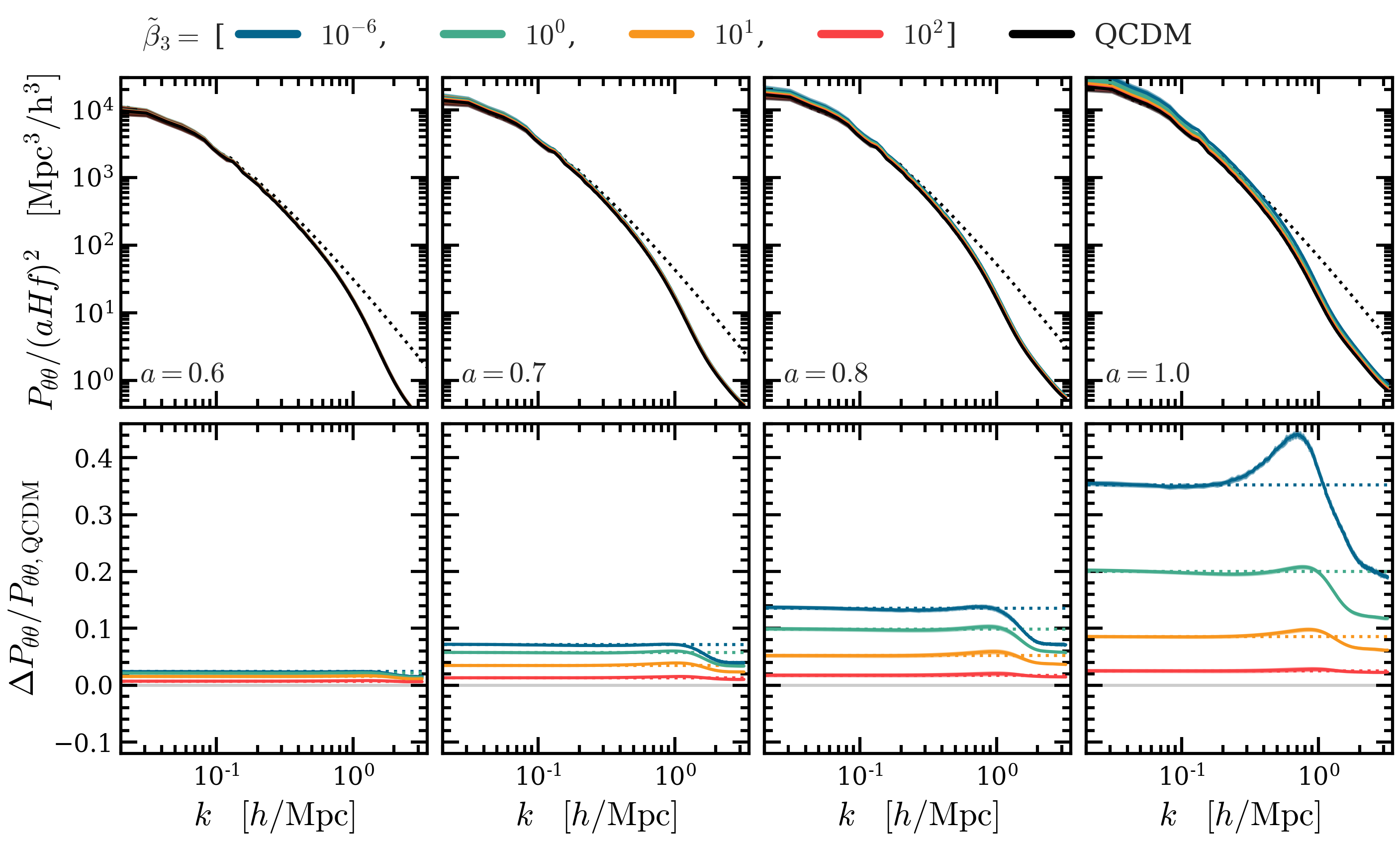}
    \caption{The velocity divergence power spectrum in the cvG model. Each column shows the results for a different scale factor: {\it outer left:} $a=0.6$, {\it inner left:} $a=0.7$, {\it inner right:} $a=0.8$, {\it outer right:} $a=1.0$. {\it Top:} The velocity divergence power spectrum of linear perturbation theory (dotted) and the cvG model for four values of $\betathree = (10^{-6}, 1, 10, 100)$, indicated by a blue, green, orange and red line respectively. {\it Bottom:} Relative differences between the velocity divergence power spectra of the cvG and QCDM models.
    }
    \label{fig:Pk_div_velo}
\end{figure}

The Kaiser formula can be improved down to quasi linear scales with additional information about the auto power spectrum of the velocity divergence\footnote{Strictly speaking, one should consider the complete velocity field, which would also involve its vorticity $\nabla_i \times v_i$. However, just as the transverse mode of the Proca field, $\nabla_i \times B_i$, it has a much smaller magnitude than its divergence and is thus neglected in SPT.}, $\theta = \nabla \cdot \mathbf{v}$, denoted as $P_{\theta\theta}$, as well as their cross spectrum $P_{\delta\theta}$, since the velocity field is more sensitive to tidal gravitational fields compared to the density field on large scales \cite{Scoccimarro:2004tg, Taruya:2010mx, Jennings:2012ej}. 

The first row of Fig.~\ref{fig:Pk_div_velo} compares the linear velocity divergence power spectrum (black dotted lines) and measured nonlinear (coloured) simulations, at $a=0.6$ (outer left), $a=0.7$ (inner left), $a=0.8$ (inner right), and $a=1.0$ (outer right). The linear power spectrum $P^{(11)}_{\theta\theta}(k;z)$ can be related to $P^{(11)}_{\delta\delta}(k;z)$ through the zeroth-moment of Eq.~\eqref{eq:vlasov}, yielding the continuity equation,
\begin{equation}\label{eq:continuity_equ}
    \dot{\delta} + \frac{1}{a}\nabla \cdot \left[\mathbf{v} \left(1+\delta\right)\right] = 0.
\end{equation}
On linear scales we can assume that the quadratic terms in Eq.~\eqref{eq:continuity_equ} vanish leaving use with
\begin{equation}\label{eq:theta_delta_relation}
    \theta = - a \dot{\delta} = - a H f \delta.
\end{equation}
Thus, the linear power spectrum of the velocity divergence is given by
\begin{equation}\label{eq:pk_linear_thetatheta}
    P^{(11)}_{\theta\theta}(k;z) = \left( a H f \right)^2 P^{(11)}_{\delta\delta}(k;z).
\end{equation}
This relation is expected to fail on non- and quasi-linear scales, as velocities grow more slowly than the linear perturbation theory predicts. Therefore, any differences in $P^{(11)}_{\theta\theta}$ between the different cvG models will appear on these scales.

In order to measure the non-linear $P_{\theta\theta}$ from the numerical simulations, we first use a Delaunay tessellation field estimator ({\tt DTFE}\footnote{The code is in the public domain, www.astro.rug.nl/~voronoi/DTFE/dtfe.html.}, \cite{Cautun:2011gf}) to obtain the volume weighted velocity divergence field on a regular grid. This procedure constructs the Delaunay tessellation from the dark matter particle locations and interpolates the field values onto a regular grid, defined by the user, by randomly sampling the field values at a given number of sample points within the Delaunay cells and then taking the average of those values. For our $500 \Mpch$ simulation boxes, we generate a grid with $512^3$ cells. From that we then measure $P_{\theta\theta}$ using the public available code {\tt nbodykit}\footnote{The code is in the public domain, nbodykit.readthedocs.io.}\cite{Hand:2017pqn}.

We can see from the top row of Fig.~\ref{fig:Pk_div_velo} that the results of the simulations for all models have approached the linear theory prediction on scales $k \lesssim 0.1 \, \hMpc$ for all times. On these scales, the time evolution of the power spectrum of all models is scale independent and, the relative difference encapsulates the modifications to the time evolution of $P^{(11)}_{\theta\theta}$ via $H$ and $f$ in Eq.~\eqref{eq:pk_linear_thetatheta}. On smaller scales, the formation of non-linear structures tends to slow down the coherent (curl-free) bulk flows that exist on larger scales. This leads to an overall suppression of the divergence of the velocity field compared to the field theory results for scales $k \gtrsim 0.1 \, \hMpc$.

A careful look into the relative difference $\Delta P_{\theta\theta}(k) / P_{\theta\theta,{\rm QCDM}}(k)$ in the bottom row of Fig.~\ref{fig:Pk_div_velo} also reveals a number of other interesting features on all scales. Firstly, we see that the wavenumber at which linear theory and simulation results for $\Delta P_{\theta\theta}(k)/P_{\theta\theta,{\rm QCDM}}(k)$ agree, $k_\ast$, depends both on $\betathree$ and the scale factor. The value of $k_\ast$ is pushed to ever larger scales as $a \to 1$ and $\betathree \to 0$. A similar observation has been made by \cite{Li:2013nua} for the DGP model. Hence, this is important for the growth rate measurement from redshift distortions. Secondly, on small scales, $k\gtrsim1\hMpc$, we can see how deviations from QCDM are suppressed by the screening mechanism, reflecting the fact that inside dark matter haloes the screening is very efficient. As also shown by $\Delta P_{\delta\delta}(k) / P_{\delta\delta,{\rm QCDM}}(k)$, the screening mechanism becomes more effective as $\betathree \to 0$. Thirdly, for $a \to 1$ and $\betathree \to 0$ we see a growing peak that for the case of $\betathree = 10^{-6}$ protrudes above the linear theory prediction at $k\sim0.7\hMpc$. A similar feature was also observed by \cite{Li:2013nua} for the DGP model.

The difference of $P_{\theta\theta}$ between the cosmological models compared to its magnitude is very small at early times, e.g., at percent level for all models when $a\lesssim0.6$, but increases rapidly over time, reaching $35\%$ for $\betathree=10^{-6}$ at $a=1.0$. This is unlike the behaviour of $\Delta P_{\delta\delta}(k) / P_{\delta\delta,{\rm QCDM}}(k)$ which increases much more slowly and only reaches $\sim5\%$ for $\betathree=10^{-6}$ at $a=1.0$. This difference is because the velocity field, being the first integration of the forces, responds more quickly to a rapid growth of the fifth-force magnitude than does the matter field, which is the second integration of the forces. It shows the rapid increase of the linear growth rate of the cvG model at late times ($a\gtrsim0.8$), and suggests that redshift-space distortions (RSD) in this time window can be a strong discriminator of this model.

\subsection{Matter bispectrum}
\label{subsec:bispectra}
As we have mentioned, even if cosmological fields are initially Gaussian, they inevitably develop non-Gaussian features as the dynamics of gravitational instability is nonlinear. Consequently, the structures found in the density field can no longer be fully described by two-point statistics alone, and higher-order correlation functions are needed in order to unlock additional information, in particular regarding the nature of gravitational interactions. To obtain first impressions of this information we use the Fourier space counterpart of the three-point correlation function, the bispectrum, which is receiving increased attention in the recent literature, not only for making more accurate predictions (see, e.g., \cite{Hashimoto:2017klo, Eggemeier:2018qae, Desjacques:2018pfv, Oddo:2019run}), but also as a probe of effects beyond \lcdm{} (e.g. \cite{GilMarin:2011xq, Bellini:2015wfa, Munshi:2016zzr, Bose:2018zpk, Bose:2019wuz, Hahn:2019zob, MoradinezhadDizgah:2020whw}).

We restrict ourselves to the study of the matter field in real space at $z=0$, for which the bispectrum is given by
\begin{equation}\label{eq:b_kkk}
    \langle \delta(\mathbf{k}_1) \delta(\mathbf{k}_2) \delta(\mathbf{k}_3) \rangle = (2\pi)^3\delta^{({\rm D})}(\mathbf{k}_1 + \mathbf{k}_2 + \mathbf{k}_3) B(\mathbf{k}_1, \mathbf{k}_2, \mathbf{k}_3),
\end{equation}
with the three wave vectors forming a closed triangle. As the study of the effects on bispectrum due to modifications to GR are still in its infancy, we shall be as comprehensive as possible by considering all possible triangle configurations between the two extreme scales $k_{\rm min}$ and $k_{\rm max}$, given a specific bin width $\Delta{k}_1=|\Delta\mathbf{k}_1|$ for each side. A detection of strong configuration dependence can be regarded as a compelling motivation to further investigate higher-order statistics. It would allow us to disentangle the modified gravity signal from other potential cosmological effects, which might be degenerate in two-point statistics and other alternative measures.

\begin{figure}
	\centering
	\includegraphics[width=.98\textwidth]{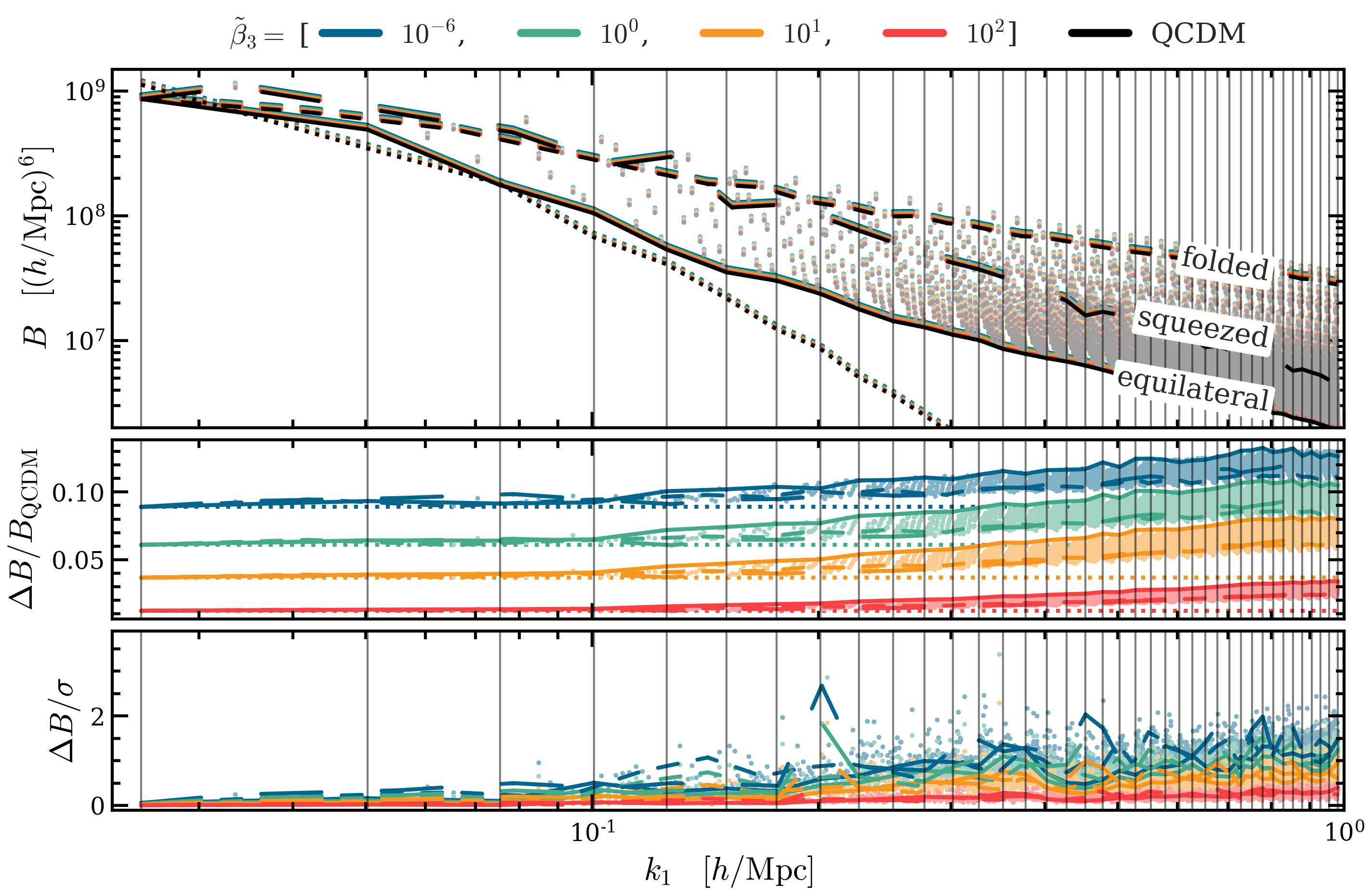}
	\caption{{\it Top}: real-space bispectrum measurements for cvG cosmologies (coloured points) and their QCDM counterpart (black points). Each data point corresponds to one of 5910 triangle configurations (see the text for more details). The vertical lines are spaced by the bin width $\Delta k \approx 0.025 \hMpc$ and indicate the value of $|\mathbf{k}_1|$, i.e., the largest triangle side. The bispectrum for equilateral configurations are shown at the tree-level (dotted), $B_{211}$, and simulation measurement (solid). The measured bispectra for the squeezed and folded configurations are shown as long and short dashed lines respectively. {\it Middle}: The relative difference between the cvG models and their QCDM counterpart. Again we show the tree-level (dotted lines) and simulation (using the same line styles as in the top panel) results. {\it Bottom}: The signal-to-noise ratio of the difference between the cvG models and their QCDM counterpart.
	}
    \label{fig:Fig_bispectrum_B}
\end{figure}

The top panel of Fig.~\ref{fig:Fig_bispectrum_B} compares the bispectrum of equilateral triangles at the tree-level (dotted line), to the measurements (solid line). It furthermore contains the measured bispectrum of squeezed triangles (long dashed), folded triangles (short dashed), and all other triangle configurations (scattered dots). Vertical lines are spaced $\Delta{k}=|\Delta\mathbf{k}|$ apart. As we assume a primordial Gaussian random field, we can apply the Wick theorem to write the bispectrum as products of power spectra summed over all possible pairings. Thus, the lowest-order bispectrum that is able to capture non-Gaussian features at late times has to expand one of the fields in the correlator of three Fourier modes to second order, yielding
\begin{equation}\label{eq:b211}
    B^{(211)}(\mathbf{k}_1, \mathbf{k}_2, \mathbf{k}_3) = \langle \delta^{(2)}(\mathbf{k}_1) \delta^{(1)}(\mathbf{k}_2) \delta^{(1)}(\mathbf{k}_3) \rangle' + \text{cyc.} = 2\mathcal{F}_2(\mathbf{k}_1, \mathbf{k}_2)P^{(11)}(k_1)P^{(11)}(k_2) + \text{cyc.},
\end{equation}
where $\delta^{(n)}$ is given in Eq.~\eqref{eq:spt_delta_order}, the primed ensemble average indicates that we have dropped the factor of $(2\pi)^3$ as well as the momentum conserving Delta function, and "cyc." stands for the two remaining permutations over $\mathbf{k}_1$, $\mathbf{k}_2$ and $\mathbf{k}_3$. Note here, that we have assumed that SPT gives an appropriate description of perturbations in the cvG model and does not fail to include further mode couplings that might be introduced through the additional Proca vector field. We will see below that this is indeed an excellent approximation. The resulting bispectrum scales as square of the linear power spectrum, $P^{(11)}$, and exhibits a strong configuration dependence as it is directly proportional to the second-order perturbation theory kernel, $\mathcal{F}_2$, which is given by,
\begin{equation}\label{eq:f2}
    \mathcal{F}_2 = \frac{17}{21} + \frac{1}{2}\frac{\mathbf{k}_1 \cdot \mathbf{k}_2}{k_1k_2} \left[\frac{k_2}{k_1} + \frac{k_1}{k_2}\right] + \frac{2}{7} \left[\frac{(\mathbf{k}_1 \cdot \mathbf{k}_2)^2}{k^2_1k^2_2} - \frac{1}{3}\right].
\end{equation}

To measure the bispectrum from the simulations, we first use fourth-order density interpolation on two interlaced cubic grids \cite{Sefusatti:2015aex} of $N=256$ cells per side. Next, we measure $B(\mathbf{k}_1,\mathbf{k}_2,\mathbf{k}_3)$ using an implementation of the bispectrum estimator presented in Ref.~\cite{Scoccimarro:2015bla}. Starting from $k_{\rm min} = 2k_f = 0.025 \hMpc$, where $k_f$ denotes the fundamental mode, we loop through all configurations satisfying $k_1 > k_2 > k_3$ and $k_1 \leq k_2 + k_3$ (the triangle closure condition). We stop after the values of $\mathbf{k}$, which are evenly spaced by $\Delta k = 2k_f$, reach the $k_{\rm max} = 1.0 \hMpc$, up until which point the shot noise is sub-dominant. With these settings -- which are chosen to keep memory consumption at bay, as it would increase rapidly otherwise -- we obtain a total of $5910$ distinct triangle configurations.

The top panel of Fig.~\ref{fig:Fig_bispectrum_B} shows that the tree-level prediction $B^{(211)}$ (dotted line) for the equilateral configuration converges to the simulation measurements of $B$ (solid line) on $k \approx 0.07 \hMpc$, which is agreement with $P_{\delta\delta}(k)$ and \cite{Baldauf:2020bsd}. In this panel we have also indicated the folded, squeezed and equilateral configurations by lines (see the legends). It does not come as a surprise that the measured bispectrum for equilateral triangles is consistently lower than all other configurations as in our considered range of $k$, the power spectrum decreases with increasing $k$ (as can be seen in Fig.~\ref{fig:Pk_matter}). The folded triangles, on the other hand, tend to have the largest amplitude, while the squeezed triangles are in between.

The middle panel of Fig.~\ref{fig:Fig_bispectrum_B} shows the relative difference, Eq.~\eqref{eq:relative_difference}, of the bispectrum of equilateral triangles at the tree-level (dotted line), and measurements (solid line); for the latter the bispectra for all triangle configurations are indicated by scattered dots. Again, the results which correspond to equilateral, squeezed and folded triangle configurations are shown by lines (the same line styles as in the top panel). We can draw the following conclusions. Firstly, as it is the case for matter and velocity divergence power spectra, the tree-level bispectrum is a good estimator on large scales ($k<k_\ast$) while the exact value of $k_\ast$ depends on redshift and the model parameter $\betathree$. However, we can see that in general linear theory gives accurate predictions of $\Delta B/B_{\rm QCDM}$ at $k < {k}_\ast \sim 0.1\hMpc$ for all models. Compared to the matter power spectra, the relative difference of the bispectra is roughly twice as large as $\Delta P_{\delta\delta} / P_{\delta\delta,\rm QCDM}$, monotonically increasing from $~1\%$ for $\betathree=100$ to $\sim 9\%$ for $\betathree=10^{-6}$. Secondly, the order of triangle configurations yielding the largest signal is reversed to the top row, with the equilateral triangles yielding the largest relative difference between cosmologies with fifth force and those without, while squeezed and folded triangles seem to converge to the same relative difference for larger values of $\betathree$. This is in agreement with \cite{Bose:2018zpk}, who arrived at a similar conclusion for $f(R)$ and DGP cosmologies. 

\begin{figure}[!h]
	\centering
	\includegraphics[width=.98\textwidth]{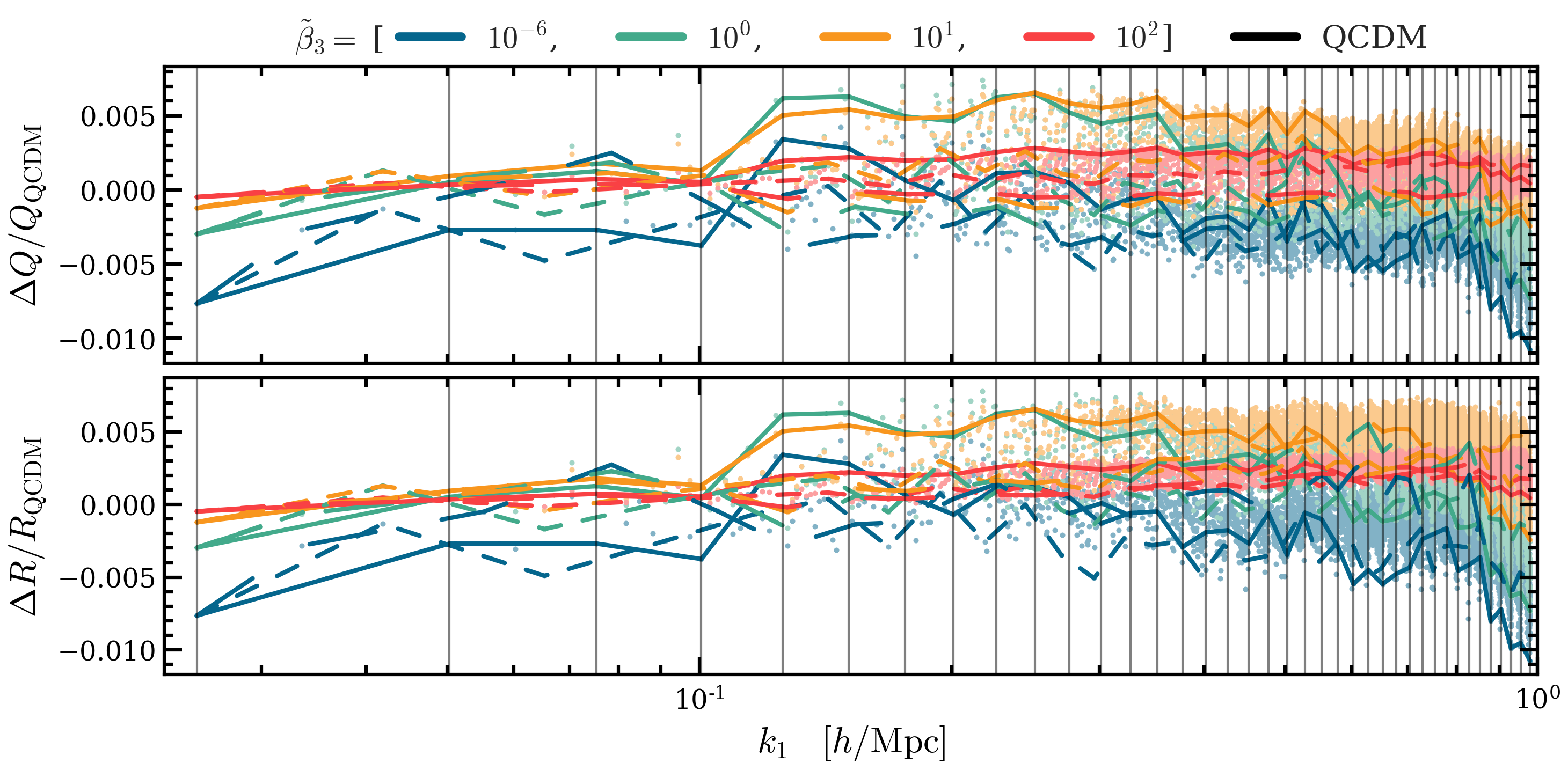}
	\caption{{\it Top}: relative difference between cvG models and their QCDM counterpart of the reduced bispectrum measurements, $Q$. {\it Bottom}: relative difference between cvG models and their QCDM counterpart of the ratio between the measured reduced bispectrum and its tree-level approximation, $Q^{(0)}$. Each data point corresponds to one of 5910 triangle configurations (see the text for more details). The lines represent equilateral (solid), squeezed (long dashed), and folded (short dashed) triangle configurations as in Fig.~\ref{fig:Fig_bispectrum_B}.
	}
    \label{fig:Fig_bispectrum_Q_R}
\end{figure}

The bottom panel of Fig.~\ref{fig:Fig_bispectrum_B} shows the SNR of the difference between cvG cosmologies and their QCDM counterpart. Three general trends are revealed: Firstly, an enhancement in the bispectrum signal with increasing $\betathree$ relative to QCDM, as we have seen in the middle panel above. Secondly, the SNR significantly increases towards smaller, nonlinear, scales. Thirdly, there is no clear trend which triangular configuration results in the highest SNR. The median taken over the range $0.1 < k \, [\hMpc] < 1$ for each cvG cosmology is: $0.88$ ($\betathree = 10^{-6}$), $0.77$ ($\betathree = 1$), $0.54$ ($\betathree = 10$) and $0.22$ ($\betathree = 100$), respectively.

A very useful statistical quantity, that isolates the configuration dependence of the triangles by removing the propagator corrections from the modified Poisson equation (contained in the nonlinear power spectrum), is the reduced bispectrum,
\begin{equation}\label{eq:q}
    Q(\mathbf{k}_1, \mathbf{k}_2, \mathbf{k}_3) \equiv \frac{B(\mathbf{k}_1, \mathbf{k}_2, \mathbf{k}_3)}{P(k_1)P(k_2) + \text{cyc.}}.
\end{equation}
The relative difference between the reduced bispectra for the cvG models and their QCDM counterpart is displayed in the top row of Fig.~\ref{fig:Fig_bispectrum_Q_R}. We indeed see how the strong scale dependencies of $\Delta B / B_{\rm QCDM}$ are removed, leaving only sub-percent deviations. The SNR of the difference of $Q$ between the cvG models and their QCDM counterpart (not shown) revealed a very weak signal on all scales for all models, with a median of $\Delta Q / \sigma \lesssim 0.05$. Therefore we shall not try to interpret the trends revealed by the individual cvG models, and instead conclude that $Q$ is very weakly dependent on $\betathree$.

To quantify how much extra mode coupling the cvG models have experienced compared to their QCDM counterpart beyond the leading term, $\mathcal{F}_2$ (defined in Eq.~\eqref{eq:f2}), we can divide the reduced bispectrum by its tree level term to define a new quantity,
\begin{equation}\label{eq:r}
    R(\mathbf{k}_1, \mathbf{k}_2, \mathbf{k}_3) \equiv \frac{Q}{Q^{(0)}} = \frac{B(\mathbf{k}_1, \mathbf{k}_2, \mathbf{k}_3)}{2\mathcal{F}_2(\mathbf{k}_1, \mathbf{k}_2)P(k_1)P(k_2) + \text{cyc.}}.
\end{equation}
The relative difference between the $R$ of the cvG models and their QCDM counterpart is displayed in the bottom row of Fig.~\ref{fig:Fig_bispectrum_Q_R}. Again, the results are in the sub-percent level and the SNR of the difference of $R$ between the cvG models and their QCDM counterpart (not shown) reveals a very weak signal on all scales for all models, with a median of $\Delta Q / \sigma \lesssim 0.06$.

The fact that for $Q$ and $R$ the relative difference between the cvG models and QCDM is fairly small, suggests that the fifth force in the cvG model does not produce substantial extra mode coupling corrections. This is a useful result because it means that the cvG effect mainly enters through the modified growth factors, which simplifies the modelling of the bispectrum.
We stress that this does not imply that the bispectrum is incapable of placing additional constraints on the cvG models. That is because the bispectrum has a different dependence on the growth factors than the power spectrum and its configuration dependence is useful in breaking degeneracies with other parameters, e.g. parameters that describe the background model or galaxy bias, such that the combination of the two statistics can still be expected to yield significant improvements.

Finally, let us note again that here we have only looked at the bispectrum of the matter density field, rather than the halo or galaxy fields. We have tried haloes, but due to the box size and resolution in our simulations, the results are noisy and the model differences unclear. Therefore we have decided not to show them here.

\section{Halo statistics}
\label{sec:halo_statistics}

This section is devoted to a detailed study of halo properties. Haloes are identified using two different algorithms, as they give complementary information about the haloes and can serve in some cases as verification. Firstly, we use the algorithm developed by \cite{Springel:2000qu} to find friends-of-friends groups to represent the ‘main' haloes, and then run \subfind{} to identify substructures in the ‘main' haloes (from now on we shall refer to the halo and subhaloes identified in this way as \subfind{} halos). Secondly, we use \rockstar{}\footnote{The code is in the public domain, https://bitbucket.org/gfcstanford/rockstar/src/main/} \cite{Behroozi:2011ju} to identify FOF haloes in the 6D phase space where substructure is more easily identifiable (from now on we will refer to these as \rockstar{} haloes). In most of this section we show results of \subfind{} haloes, although we have checked that the {\tt ROCKSTAR} haloes give similar results. We use \rockstar{} haloes to study the halo concentration mass relation, because this is directly measured by \rockstar{}.

Note that, in principle, the unbinding procedure employed by the halo finding algorithms would need to be modified due to the presence of the fifth force induced by the Proca field. However, \cite{Li:2010mqa} found the effect of this modification to be quite small for chameleon models. Also, we will see below, the fifth force in the cvG models is strongly suppressed by Vainshtein screening, and so we expect its effect will be even smaller here. Thus, we use identical versions of \subfind{} and \rockstar{} for the different cosmologies.

We compare the cvG models to their QCDM counterpart in the same way as we have done in Sec.~\ref{sec:matter_field_statistics} via Eq.~\eqref{eq:relative_difference} and Eq.~\eqref{eq:snr}.

\subsection{Halo mass function}
\label{subsec:halo_mass_functions}

We start the analysis of the halo populations with the one-point distribution of halo masses -- the halo mass function (HMF). The halo mass is defined as the mass enclosed in the spherical region of radius $R_{200}$ around the centre of the over-density, within which the mean density is $200$ times the critical density $\rho_c$ at the halo redshift,
\begin{equation}\label{eq:m200}
    M_{200c} = \frac{4\pi}{3}R_{200}^3200\rho_c, \, \text{ with } \, \rho_c = \frac{3H^3}{8\pi G}.
\end{equation}

\begin{figure}[!h]
	\centering
	\includegraphics[width=.98\textwidth]{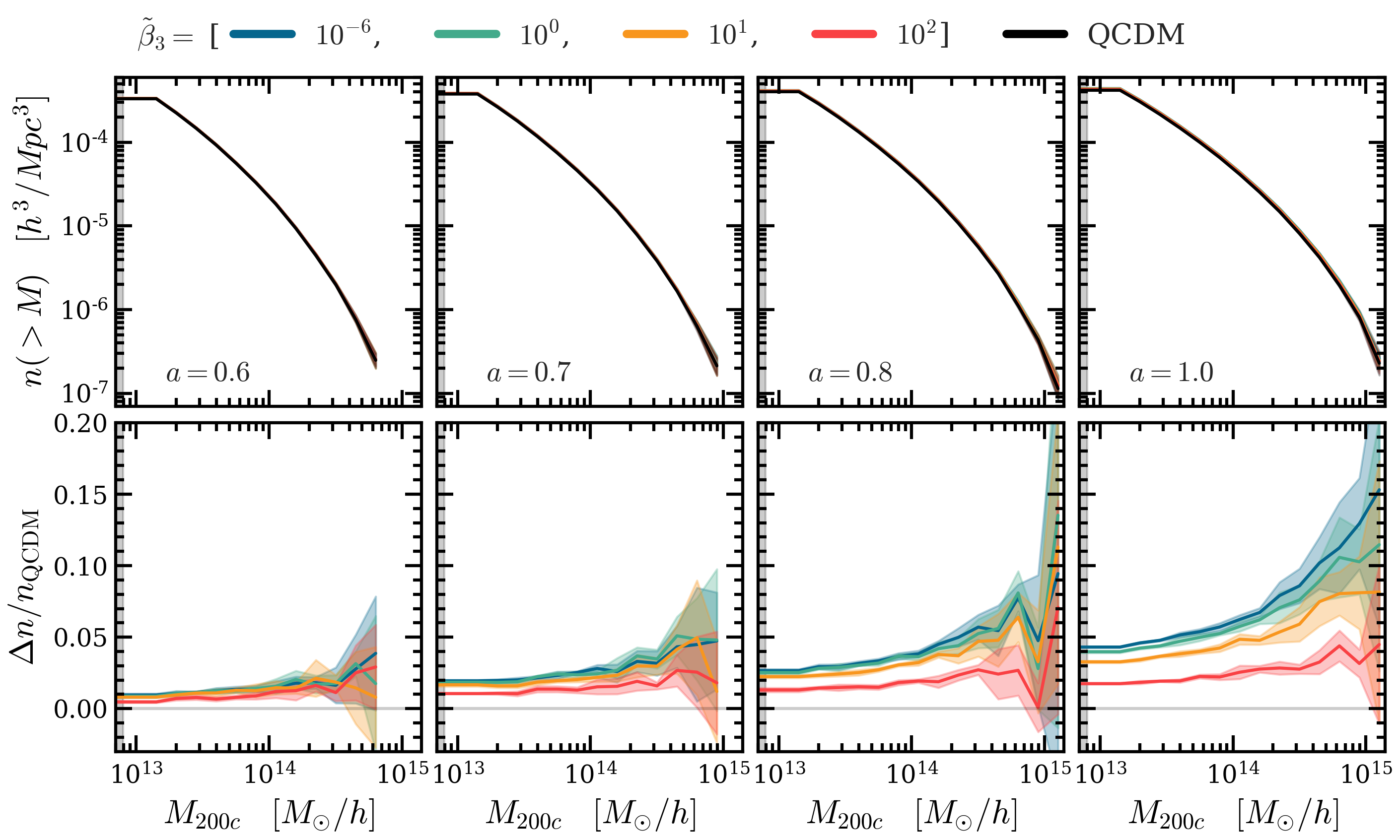}
	\caption{{\it Top}: panels show the cumulative halo mass function, $n\left(>M_{200c}\right)$, for the cvG model (coloured) and their QCDM (black) counterpart. Each column shows the results for a different scale factor: {\it outer left}: $a=0.6$, {\it inner left}: $a=0.7$, {\it inner right}: $a=0.8$, {\it outer right}: $a=1.0$. {\it Bottom}: the relative differences to QCDM. The results shown are obtained by averaging over the simulations of the 5 different initial condition realizations and the shaded region show the standard deviation over these realizations. The vertical shaded region corresponds to haloes with fewer than $100$ simulation particles, for which the number is incomplete due to the lack of resolution.
	}
    \label{fig:halo_mass_fct}
\end{figure}

In the top row of Fig.~\ref{fig:halo_mass_fct} we show the cumulative HMF, $n(>M_{200c})$, which is the number density of dark matter haloes more massive than the given $M_{200c}$, at $a=0.6$ (outer left), $0.7$ (inner left), $0.8$ (inner right) and $1.0$ (outer right). The {\it bottom-up} picture of structure formation, i.e., small-scale objects collapse first and merge to form increasingly massive objects as time proceeds, is clearly visible, which follows from the fact that in our model dark matter is cold.

The bottom row of Fig.~\ref{fig:halo_mass_fct} shows the relative difference between the cvG models and their QCDM counterpart. The median of SNR of the differences between the models over the range shown in the figure is: $7.1$ ($\betathree = 10^{-6}$), $6.4$ ($\betathree = 1$), $5.5$ ($\betathree = 10$), $2.9$ ($\betathree = 100$). We find good agreement with \cite{Barreira:2013eea}, and have verified that the result is consistent between \subfind{} and \rockstar{}. The fifth force enhances the abundance of dark matter haloes in the entire mass range probed by the simulations, with the enhancement stronger at late times and for high-mass haloes, which mimics the effect of the csG model \cite{Barreira:2014zza}. This is to be expected because the strength of the fifth force increases over time \cite{Becker:2020azq}. Note that for massive haloes the increase in abundance is mainly due to an increase in individual halo masses, as can be seen from the top panels: we remark that more massive haloes are not necessarily more strongly screened in Vainshtein models (see, e.g., Fig.~8 of \cite{Hernandez-Aguayo:2020kgq}), and the enhanced gravity around these massive haloes helps to bring more matter from their (matter-rich) surroundings to their vicinity, allowing them to grow larger. On the other hand, models with more efficient screening, such as $\betathree>1$, show a more restrained enhancement of the HMF.

\subsection{Two-point correlation functions}
\label{subsec:correlation_functions}

The configuration-space counterpart of the matter power spectrum, $P_{\delta\delta}$, presented in Sec.~\ref{subsec:power_spectra}, is the two-point correlation function (2PCF), $\xi(r)$. In principle these two measures would carry the same information, but in practice this is not guaranteed since our analyses are restricted to a finite range of scales, and moreover, configuration and Fourier space statistics are impacted by different systematic effects.

\begin{figure}[!h]
	\centering
	\includegraphics[width=.98\textwidth]{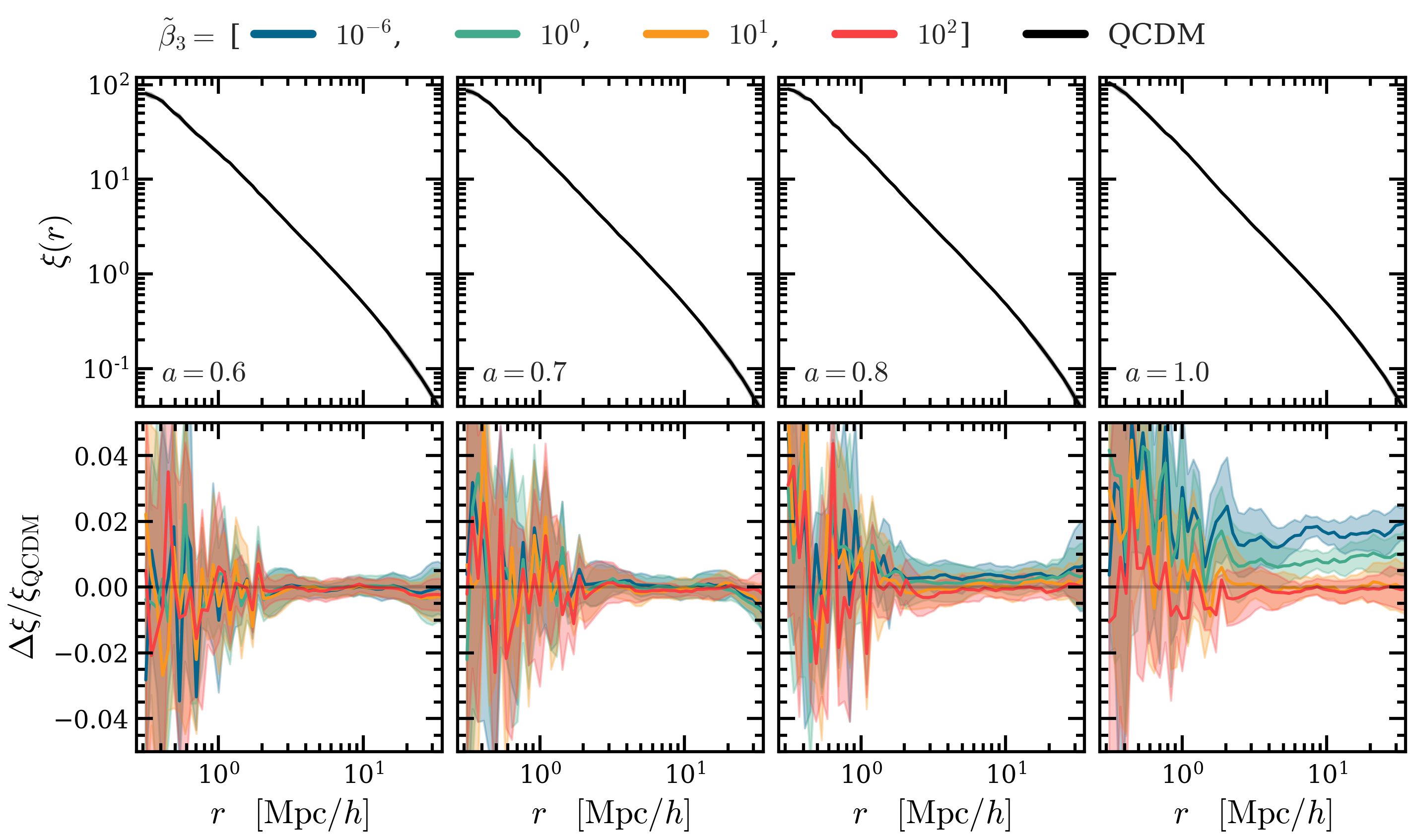}
	\caption{{\it Top}: The halo 2PCFs in the QCDM model. Each column shows the results for a different scale factor: {\it outer left}: $a=0.6$, {\it inner left}: $a=0.7$, {\it inner right}: $a=0.8$, {\it outer right}: $a=1.0$. Note that to prevent the plot from appearing cluttered we have not shown the results for the cvG models. Furthermore, we included the standard deviation as a shaded region, but it is too small to see. {\it Bottom}: The relative differences between models. The cvG model for four values of $\betathree = (10^{-6}, 1, 10, 100)$ are shown, indicated by a blue, green, orange and red line respectively. The shaded regions are the standard deviations among the five simulation realizations.
	}
    \label{fig:Fig_two_point_corr_fct}
\end{figure}

For this analysis we use \subfind{} haloes, since these catalogues contain the subhaloes which can be proxies of satellite galaxies, and without which $\xi(r)$ would decay at $r\lesssim1$-$2\Mpch$ due to the halo exclusion effect. We show their respective 2PCFs in the top row of Fig.~\ref{fig:Fig_two_point_corr_fct} for $a=0.6$ (outer left), $a=0.7$ (inner left), $a=0.8$ (inner right) and $a=1.0$ (outer right). As expected, the 2PCFs drop off with halo separation, and can be well described by a power law across the entire range of scales probed here.

The relative difference between the 2PCFs of the cvG models and their QCDM counterpart for \subfind{} haloes is shown in the bottom row of Fig.~\ref{fig:Fig_two_point_corr_fct}. As for the power spectrum of the matter field, Fig.~\ref{fig:Pk_matter}, we see more enhanced clustering for smaller values of $\betathree$. However, the cvG enhancement for halo clustering is smaller than for matter clustering, implying slightly smaller halo biases in stronger cvG models. This is because haloes are biased tracers of the dark matter field, and their bias generally decreases over time, as structure formation progresses: the enhanced gravity in cvG models simply speeds this up. Note that the enhancement of the halo 2PCF is nearly constant down to $\sim3\Mpch$, consistent with the behaviour of the matter power spectrum (cf.~Fig.~\ref{fig:Pk_matter}), and reflecting the fact that in the cvG model the growth factor is enhanced in a scale-independent way in the linear regime.

\subsection{Mean halo pairwise velocity}
\label{subsec:mean_pairwise_velocity}

As outlined earlier, it is quintessential to develop a theoretical model of the pairwise velocity statistics as well as the real-space correlation function for cosmological analyses with redshift surveys, such as Euclid and DESI. Although we do not strive to actually test the cosmological models investigated here, we measure the relevant quantities to gain an intuition of how they are affected by the cvG model and to aid future work.

\begin{figure}[!h]
	\centering
	\includegraphics[width=.98\textwidth]{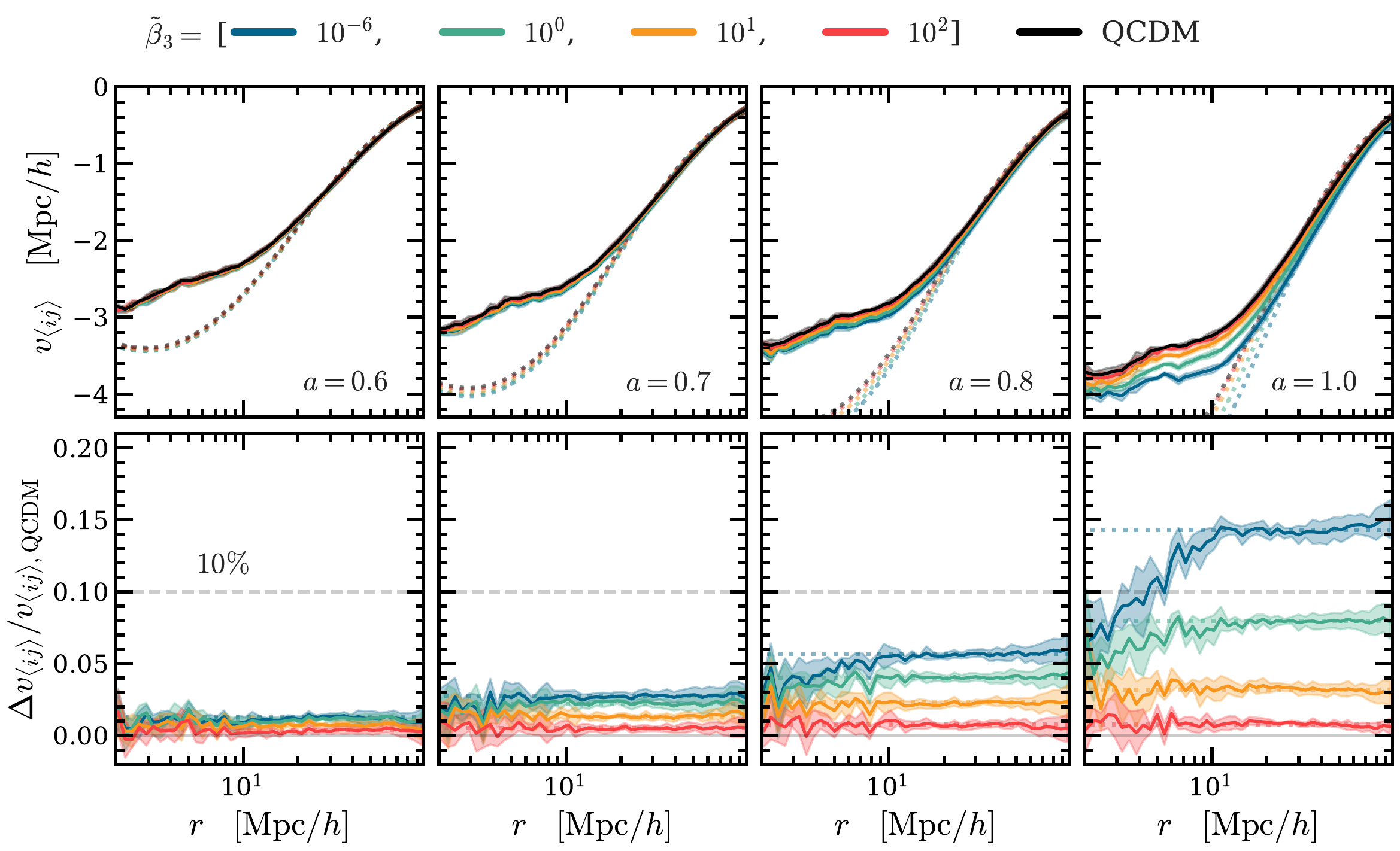}
	\caption{{\it Top}: the mean pairwise radial velocity of dark matter haloes. In each panel we show the mean measurements from the simulations (solid lines) with their one standard deviation (shaded regions), together with the linear theory predictions (dotted lines). Each column shows the results for a different scale factor: {\it outer left}: $a=0.6$, {\it inner left}: $a=0.7$, {\it inner right}: $a=0.8$, {\it outer right}: $a=1.0$. {\it Bottom}: the relative differences between the cvG models and QCDM. Note that the velocities are rescaled by $H$ so that they have the unit of length.
	}
    \label{fig:Fig_pairwise_velocity_00}
\end{figure}

For this analysis we use \subfind{} haloes, as they contain the smallest haloes and subhaloes and thus can enable measurements to smaller scales, including the virial motions of subhaloes inside main haloes. We show the measured mean pairwise velocities for the different models in the top row of Fig.~\ref{fig:Fig_pairwise_velocity_00}, comparing linear estimates (dotted lines) to the simulation results (solid lines) at $a=0.6$ (outer left), $a=0.7$ (inner left), $a=0.8$ (inner right) and $a=1.0$ (outer right). The linear mean pairwise velocity, $v_{\langle ij \rangle}$, is intimately related to the 2PCF of the matter field, $\xi(r)$, through the pair conservation equation, Eq.~\eqref{eq:pairwise_continuity_equ}, just as $P_{\theta\theta}$ is to $P_{\delta\delta}$ (see Sec.~\ref{subsec:power_spectra}) through the continuity equation, Eq.~\eqref{eq:continuity_equ} \cite{peebles1980large},
\begin{equation}\label{eq:pairwise_continuity_equ}
    \dot{\xi}_{ij} + \frac{1}{a}\nabla_{ij} \cdot \left[v_{\langle{ij}\rangle} \left(1+\xi_{ij}\right)\right] = 0.
\end{equation}
We can replace the 2PCF in Eq.~\eqref{eq:pairwise_continuity_equ} with its Fourier space counterpart in first order, $P^{(11)}_{\delta\delta}$, using the first-order Bessel function $j_1$, and obtain the linear theory prediction of $v_{\langle ij \rangle}$ expressed as
\begin{equation}\label{eq:v12_linear}
    v_{\langle ij \rangle}(r) = -r \frac{fb}{\pi^2} \int dk P^{(11)}_{\delta\delta}(k) j_1(kr) k ,
\end{equation}
where $b$ is the linear bias of halos, $f$ is the linear growth rate and $j_1$ is the spherical Bessel function of order $1$ \cite{Reid:2011ar}. To get the bias values used in the linear theory prediction for Fig.~\ref{fig:Fig_pairwise_velocity_00}, cf.~Eq.~\eqref{eq:v12_linear}, we compute the halo power spectrum, $P_{hh}$, divide it by the matter power spectrum, $b^2 \approx P_{hh} / P_{\delta\delta}$. Due to the sparseness of haloes, the shot-noise becomes sub-Poisson on larger scales than it does for dark matter particles. Therefore we restrict the calculation of $b$ to scales where the relation stays approximately constant, $0.025 < k \, \Mpch < 0.1$. We find that at each scale factor, the different cosmological models have the same fitted value of $b$ (averaged over all 5 simulation realisations) up to the second decimal. Beyond the second decimal $b$ indeed increases with $\betathree$ as expected from the relation of $\xi(r)$ and $P_{\delta\delta}(k)$.

The relative difference between $v_{\langle ij \rangle}$ of the cvG models and their QCDM counterpart is shown in the bottom row of Fig.~\ref{fig:Fig_two_point_corr_fct}, which seems to have converged to a constant value for all cvG models at scales $r > 10 \, \Mpch$.  As an example, for $\betathree=10^{-6}$ the relative difference settles on $\sim 0.15$ for large scales, which is approximately half of $\Delta P_{\theta\theta}(k) / P_{\theta\theta,{\rm QCDM}}(k)$ shown in Fig.~\ref{fig:Pk_div_velo}, partially due to the fact that $P_{\theta\theta} \propto f^2$. If \rockstar-halos are considered the same qualitative trend is found on the larger scales.

\subsection{Redshift space clustering}
\label{subsec:redshift_space_clustering}

Motivated by the results of the real space clustering and mean pairwise velocity, we carry on to study the halo 2PCF in redshift space. In real observations, instead of their radial distances, we measure the redshifts of galaxies. The conversion from redshift space to real-space galaxy coordinates is not only determined by the Hubble expansion, but also affected by the peculiar velocities of galaxies. This induces anisotropies on what would be an isotropic galaxy correlation function, known as redshift-space distortions (RSD). RSD is a useful probe of the peculiar velocity field, and consequently the growth rate of matter. In particular, the quadrupole of the redshift-space galaxy correlation function is sensitive to the galaxy (or halo) pairwise infall velocity, which we have seen above can be strongly enhanced by the fifth force in the cvG model. We use haloes (subhaloes) as proxies of galaxies in this study.

\begin{figure}[!h]
	\centering
	\includegraphics[width=.98\textwidth]{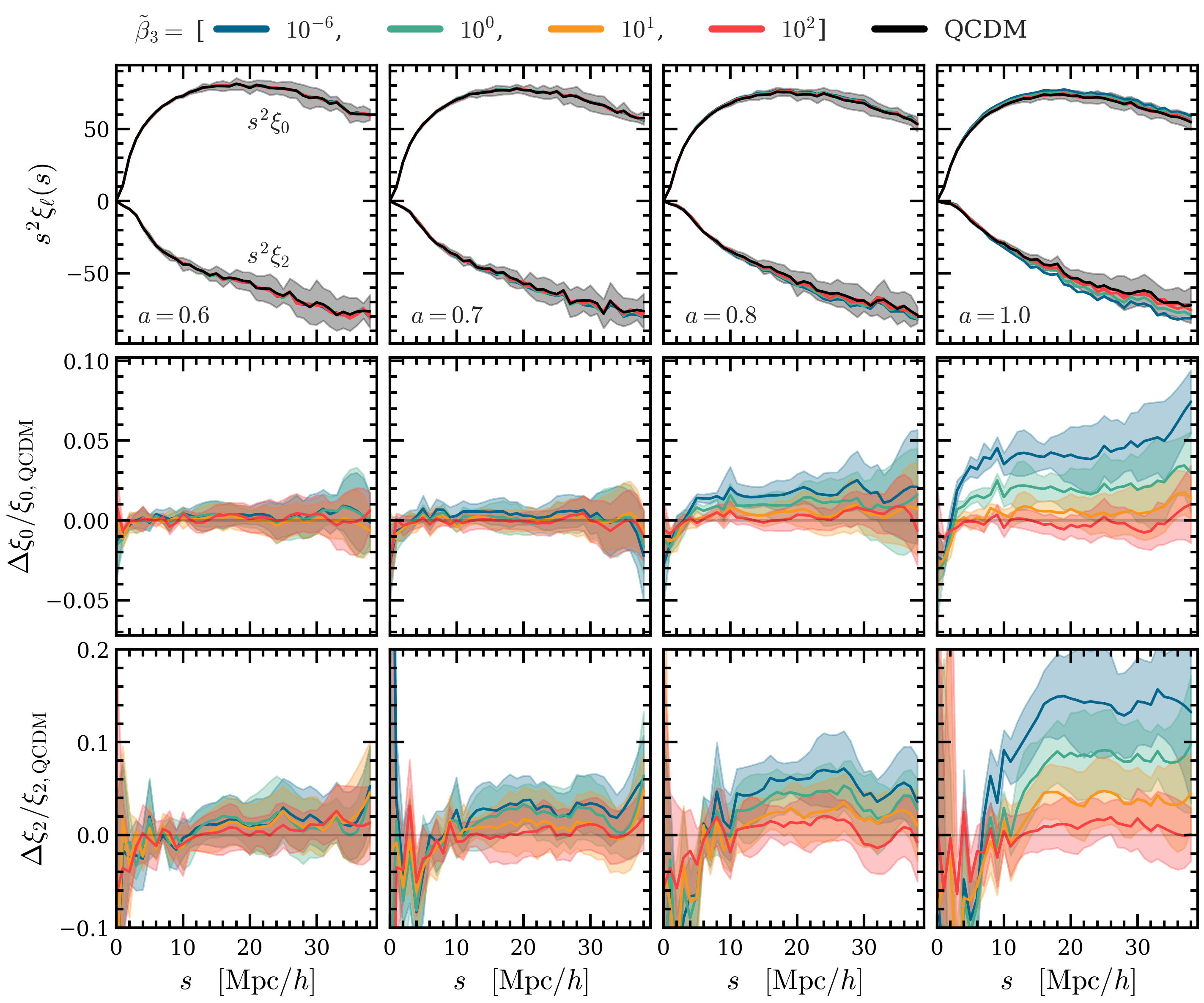}
	\caption{{\it Top}: the monopole, $\xi^{\rm s}_{0}$, and quadrupole, $\xi^{\rm s}_{2}$, moments of the 2PCF in redshift space. The results are obtained by averaging over the five simulations for each cosmology (solid lines) and shaded region show the standard deviation over these realization, which we show only for QCDM to maintain clearness. We have not shown the cvG results to prevent the plot from appearing cluttered. {\it Central and bottom}: the relative differences of $\xi^{\rm s}_{0}$ and $\xi^{\rm s}_{2}$ respectively. Each column shows the results for a different scale factor: {\it outer left}: $a=0.6$, {\it inner left}: $a=0.7$, {\it inner right}: $a=0.8$, {\it outer right}: $a=1.0$.
	}
    \label{fig:Fig_tpcf_s}
\end{figure}

The mapping of the halo coordinates from real space to redshift space is given by,
\begin{equation}
    \mathbf{s} = \mathbf{r} + \frac{\mathbf{v}(\mathbf{r})\cdot \hat{z}}{a\mathcal{H}} \hat{z},
\end{equation}
where $\hat{z}$ is the unit vector in the line of sight direction which we have chosen to be along the $z$-axis of the simulation box, assuming that the galaxies are far away from the observer ({\it plane-parallel} approximation). Thus, the anisotropic correlation function is given by
\begin{equation}
    \xi^{\rm s}(s, \mu) = \langle \delta(\mathbf{x}) \delta(\mathbf{x} + \mathbf{s}) \rangle,
\end{equation}
where ${\bf s}$ is the halo separation vector, $s$ its magnitude, $s_\parallel$ the halo separation along the line of sight direction, and $\mu = \cos(s_{\parallel} / s)$ is the cosine of the angle between $\mathbf{s}$ and the LOS. We measure $\xi^{\rm s}(s, \mu)$, using \subfind-halos for the same reason stated in the previous section, over $40$ bins of $\mu = [0, 1]$ and $40$ bins of $s = [0, 40] \, \Mpch$. In order to increase the SNR ratio, it is helpful to project $\xi^{\rm s}(s, \mu)$ onto a one-dimensional object which depends on $s$ only. Therefore, we decompose the measured $\xi^{\rm s}(s, \mu)$ into multipole moments using its Legendre expansion,
\begin{equation}\label{eq:2pcf_moments}
    \xi^{\rm s}(s, \mu) = \sum_{\ell} \xi^{\rm s}_{\ell}(\mu) L_{\ell}(\mu),
\end{equation}
where $\ell$ is the order of the multipole and $L_{\ell}(\mu)$ is the Legendre polynomial at the $\ell$-th order. Inverting Eq.~\eqref{eq:2pcf_moments} and integrating over $\mu$, we find
\begin{equation}\label{eq:2pcf_inv_moments}
    \xi^{\rm s}_{\ell}(s) = \frac{2\ell + 1}{2} \int^{1}_{-1}{\rm d}\mu\xi^{\rm s}(s, \mu)L_{\ell}(\mu).
\end{equation}
As the redshift space correlation function is symmetric in $\mu$, only even values of $\ell$ give a non-zero contributions. Of these, we study the two lowest multipoles: the monopole ($\ell = 0$), and the quadrupole ($\ell = 2$). We omit higher order multipoles ($l\geq4$), as they do not have a big impact on the estimation of the correlation function and are noisier than the monopole and quadrupole \cite{Hamilton:1998lrd}.

In the top row of Fig.~\ref{fig:Fig_tpcf_s}, we show the monopole, $\xi^{\rm s}_{0}$, and quadrupole, $\xi^{\rm s}_{2}$, moments of the QCDM model, at $a=0.6$ (outer left), $0.7$ (inner left), $0.8$ (inner right) and $1.0$ (outer right). We limit the study to scales $< 40 \Mpch$ which is roughly $1/10$ of the simulation box size. We know, however, that the peak position of the baryon acoustic oscillations (BAO) will be affected by the cvG model, as $\betathree \to \infty$ converges to QCDM and $\betathree \to 0$ converges to the cosmology of the csG, both being different from \lcdm{}. The csG model is known to be unable to reproduce the BAO position \cite{deFelice:2017paw,Nakamura:2018oyy,DeFelice:2020sdq} (see however \cite{Barreira:2014ija}).

The central and bottom rows of Fig.~\ref{fig:Fig_tpcf_s} show the relative differences between the cvG models and their QCDM counterpart, for the monopole and quadrupole, respectively. The quadrupole moment encodes the anisotropies induced by redshift distortions, and as it has been the case for $\xi$ and $v_{\langle ij \rangle}$, the relative difference of the cvG model to its QCDM counterpart increases with a decreasing value of $\betathree$ especially on scales $>20 \Mpch$. This implies that with decreasing $\betathree$ the contours of the two-dimensional 2PCF in redshift space, $\xi^{\rm s}(s_{\parallel}, s_{\perp})$, are more squashed, which is a direct consequence of the enhanced growth rate and stronger matter fluctuations as could already be anticipated from the results shown in Fig.~\ref{fig:Pk_matter}. The values of $\Delta \xi_2 / \xi_{2, {\rm QCDM}}$ converge on large scales for each cvG model to approximately the same values as for $\Delta{v}_{\langle ij \rangle} / {v}_{\langle ij \rangle, {\rm QCDM}}$. The median SNR at $a=1$ (outer right panel), taken over the range $20 < s/(\Mpch) < 40$, is approximately equal up to $7.2$ for the monopole and $3.5$ for the quadrupole for the strongest cvG model $\betathree=10^{-6}$. Although the relative difference is larger in the quadrupole, the SNR values are larger for the monopole, which is because the quadrupole is sensitive to the pairwise infall velocity $v_{\langle{ij}\rangle}$, which has a larger scatter than the real-space correlation function (see Figs.~\ref{fig:Fig_pairwise_velocity_00} and \ref{fig:Fig_two_point_corr_fct}) that dominates the monopole signal. The RSD quadrupole can be a more promising probe to constrain the cvG model if the statistical uncertainties can be reduced by large amount of data.

\subsection{Concentration-mass relation}
\label{subsec:mass_concentration_relation}

For dark matter haloes, the strongest effect of Vainshtein screening is perhaps in the density profiles. This is because the interiors of haloes are expected to be strongly screened, see e.g., \cite{Falck:2014jwa,Falck:2015rsa,Hernandez-Aguayo:2020kgq}. The Vainshtein screening radius can be even larger in the csG model and cvG models with $\betathree\rightarrow0$, than in the DGP model at late times \cite{Becker:2020azq}, so we expect the screening to be strong and the internal properties of haloes protected by it from the influence of the fifth force.

The density distribution inside dark matter halos is well described by the universal Navarro-Frenk-White (NFW; \cite{Navarro:1995iw,Navarro:1996gj}) profile,
\begin{equation}\label{eq:nfw}
    \rho_{\rm NFW}(r) = \frac{\rho_s}{r/R_s \left(1 + r/R_s\right)^2},
\end{equation}
where $\rho_s$ and $R_s$ are the characteristic density and scale radius respectively, which can vary from halo to halo. Thus the halo mass, $M_{200c}$, can be obtained by integrating the NFW density profile
\begin{equation}\label{eq:nfw_m200_c200_rel}
    M_{200c} = \int_0^{R_{200c}} dr 4\pi \rho_s^2 \rho_{\rm NFW}(r) = 4\pi r^2 \frac{R_{200c}^3}{c_{\rm 200}^3} f\left(c_{200}\right),
\end{equation}
where we have defined the function
\begin{equation}
    f(x) = {\rm ln}\left(1 + x\right) - \frac{x}{1+x},
\end{equation}
and the concentration parameter, 
\begin{equation}\label{eq:nfw_c200}
    c_{\rm 200} \equiv \frac{R_{200c}}{R_s},
\end{equation}
which describes the steepness of the density profile. Using Eq.~\eqref{eq:nfw}, we can relate $\rho_s$ to $c_{200}$, and therefore the NFW profile can be fully parametrised using $M_{200}$ and $c_{200}$. Here we use the publicly available phase-space friends-and-friends code \rockstar{} \cite{Behroozi:2011ju} to calculate the halo concentrations. \rockstar{} solves the concentration using the following equation:
\begin{equation}
    \frac{GM_{200c}}{R_{200c}}\frac{c_{200}}{f\left(c_{200}\right)} = v^2_{\rm max}\frac{2.163}{f(2.163)},
\end{equation}
where $v_{\rm max}=\sqrt{GM(<R_{\rm max})/R_{\rm max}}$ is the maximum circular velocity inside a halo, which occurs at $r=R_{\rm max}\simeq2.163R_s$ for an NFW density profile. Note that we do not attempt to do a full fitting of the NFW profile Eq.~\eqref{eq:nfw} for individual haloes in this work.

\begin{figure}[!h]
	\centering
	\includegraphics[width=.98\textwidth]{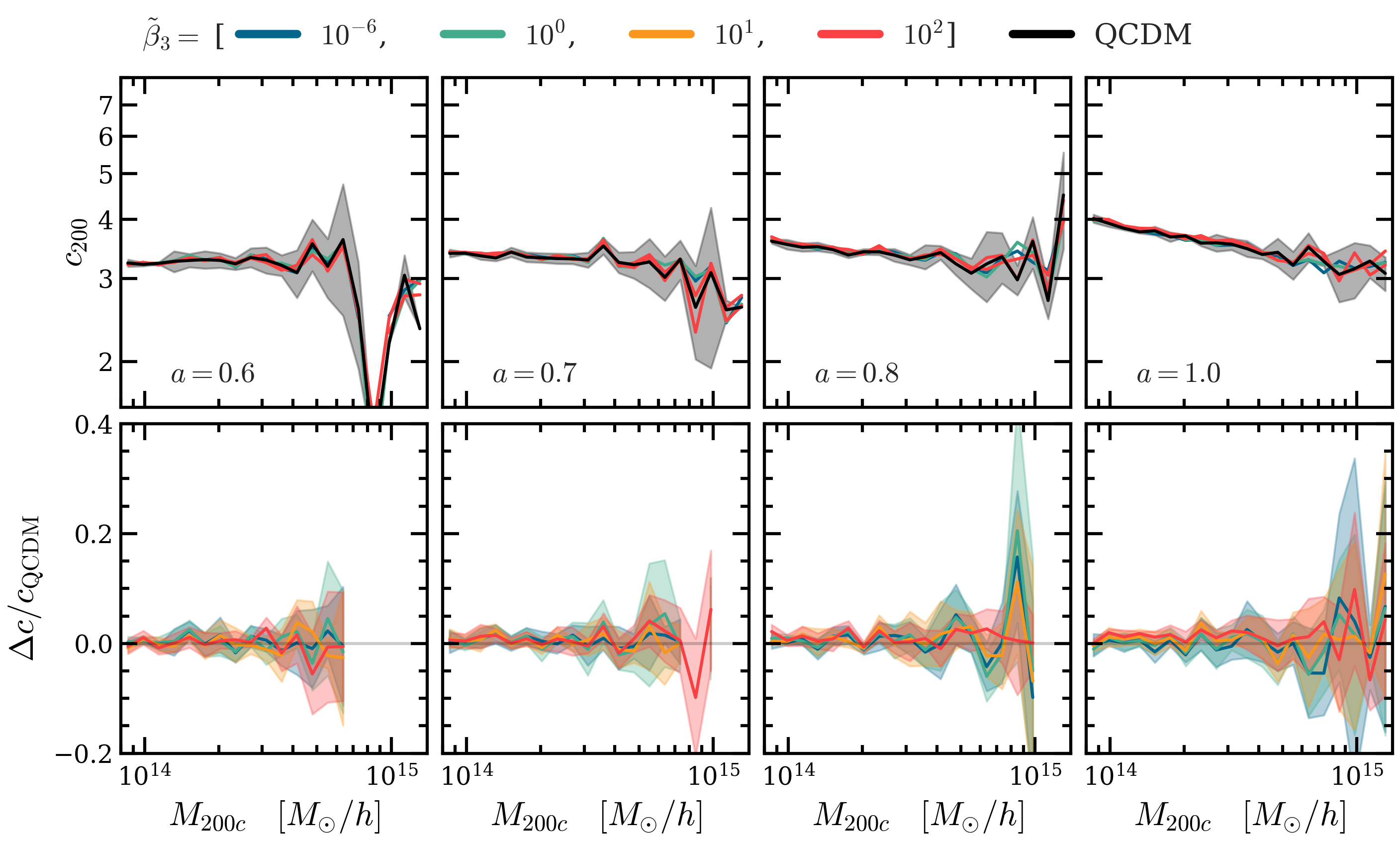}
	\caption{The top panels show the relationship between halo mass, $M_{200c}$, and the NFW definition of halo concentration, $c_{\rm 200}$ for the QCDM model (the results for the cvG variants are not shown here because they are very close to the QCDM one). The bottom panels show the relative differences of the cvG models to QCDM, $\Delta c/c_{\rm QCDM} = \left(c_{\rm cvG} - c_{\rm QCDM}\right)/c_{\rm QCDM}$. Each column shows the results for a different scale factor: {\it outer left}: $a=0.6$, {\it inner left}: $a=0.7$, {\it inner right}: $a=0.8$, {\it outer right}: $a=1.0$. The results shown are obtained by averaging over the 5 independent realisations of simulations, and the shaded region show the standard deviation over these realisations. We do not show the results for $M_{200} \gtrsim 1.5\times10^{15}M_{\odot}/h$, since in this mass ranges there are only a few haloes.
	}
    \label{fig:concentration_mass_rel}
\end{figure}

The top row of Fig.~\ref{fig:concentration_mass_rel} shows the halo concentration-mass relations at $a=0.6$ (outer left), $a=0.7$ (inner left), $a=0.8$ (inner right), and $a=1.0$ (outer right). To ensure accurate measurements, we have excluded all haloes with fewer than 1000 simulation particles from this figure which, combined with the small box size of our simulations, allows us to analyse the $c_{200}$-$M_{200c}$ relationship for halo masses that span only one order of magnitude. Nevertheless, we can clearly see that the relationship follows a power law  \cite{Maccio:2006wpz,Neto:2007vq,Duffy:2008pz}. Note that the statistics is poor at large mass and early times, due to a lack of haloes.

Without the screening mechanism we would expect haloes in a Proca universe to be more concentrated than their counterparts in a QCDM cosmology, since the strength of gravity increases quickly at late times \cite{Becker:2020azq}, which causes a faster steepening of the gravitational potential inside haloes, attracting more matter to the central region and leading to a steeper density profile \cite{Mitchell:2019qke}. However, in the cvG model in reality, just as for the csG model \cite{Barreira:2014zza}, inside haloes the Vainshtein screening is strong enough that there is little effect of the fifth force, as can be seen from the bottom panels of Fig.~\ref{fig:concentration_mass_rel}.

\section{Weak Lensing statistics}
\label{sec:lensing_statistics}

In the final section we focus on the study of weak-lensing statistics. We start by analysing the lensing convergence field (\orig) which can be used together with the matter power spectrum and bispectrum to circumvent the dependence on tracer bias (e.g., \cite{Zhang:2007nk}), and end with an analysis of the abundances and tangential shear profiles of voids identified from WL maps \cite{Davies:2018jpm,Davies:2020xem}.

\subsection{Weak lensing convergence and peak statistics}
\label{subsec:weak_lensing}

Weak lensing (WL) is governed by the lensing potential, $\Phi_{\rm lens}$, which is given by
\begin{equation}\label{eq:phi_lens}
    \Phi_{\rm lens} = \frac{\Phi+\Psi}{2},
\end{equation}
with $\Phi$ and $\Psi$ being the two Bardeen potentials in the metric Eq.~\eqref{eq:FLRW_metric}. $\Phi$ and $\Psi$ are related to each other through the anistropic stress. At late times, since we neglect matter species such as photons and neutrinos, in the cvG and QCDM models, the anisotropic stress is negligible so that we have $\Phi=\Psi$. Therefore, in the cvG model not only massive particles can feel deviations from GR, but also can massless particles, as the dynamical and lensing potentials are equal and can both be modified substantially in the case of $\betathree\rightarrow0$. This is in contrast to some other models of gravity, such as $f(R)$ gravity and the DGP model.

The relation between \orig{} and $\Phi_{\rm lens}$ and how those quantities are solved ‘on-the-fly' during the simulation run time was summarised in Sec.~\ref{sec:simulations_section}. Here we would like to be more explicit how Eq.~\eqref{eq:kappa} is affected by the cvG compared to the QCDM. For the QCDM cosmology we have, in \ecosmog's code units,
\begin{equation}
    \tilde{\nabla}^2 \tilde{\Phi}^{\rm QCDM}_{\rm lens} = \tilde{\nabla}^2 \tilde{\Phi}^{\Lambda{\rm CDM}}_{\rm lens} = 4\pi Ga^2 \delta\tilde{\rho},
\end{equation}
where $G$ is the gravitational constant and $\delta\tilde{\rho}$ the density contrast. However, as the expansion history is altered in QCDM compared to \lcdm{} their $\kappa$ field will not be the same. For the cvG model, where the fifth force and screening mechanism are included, the lensing potential is
\begin{equation}
    \tilde{\nabla}^2 \tilde{\Phi}^{\rm cvG}_{\rm lens} = \tilde{\nabla}^2 \tilde{\Phi}^{\Lambda{\rm CDM}}_{\rm lens} + \frac{3\beta_{\rm sDGP}}{2\beta}\alpha\tilde{\partial}^2\tilde{\chi},
\end{equation}
where $\beta_{\rm sDGP}$ is the coupling strength between matter and the brane-bending mode in the sDGP model, and $\beta$ and $\alpha$ are given by Eq.~\eqref{eq:func_beta} and Eq.~\eqref{eq:func_alpha} respectively. This modification of the lensing potential will modify Eq.~\eqref{eq:kappa} in the linear regime as
\begin{equation}\label{eq:kappa_cvg}
    \kappa = \frac{1}{c^2}\int_0^{\chi_s}\frac{\chi\left(\chi_s - \chi\right)}{\chi_s} \left(1 + \frac{\alpha}{\beta}\right) {\tilde{\nabla}}^2\tilde{\Phi}_{\rm lens, 2{\rm D}}(\chi, \vec{\beta}(\chi)) {\rm d}\chi,
\end{equation}
in addition to the modified expansion history. Here $\chi$, which is the comoving distance, should not be confused with the longitudinal Proca mode, $\tilde{\chi}$. This simple rescaling does not account for the effects of the screening mechanism and can only be accurately predicted through simulations as used in this work.

\begin{figure}[!b]
    \centering 
    \includegraphics[width=.98\textwidth]{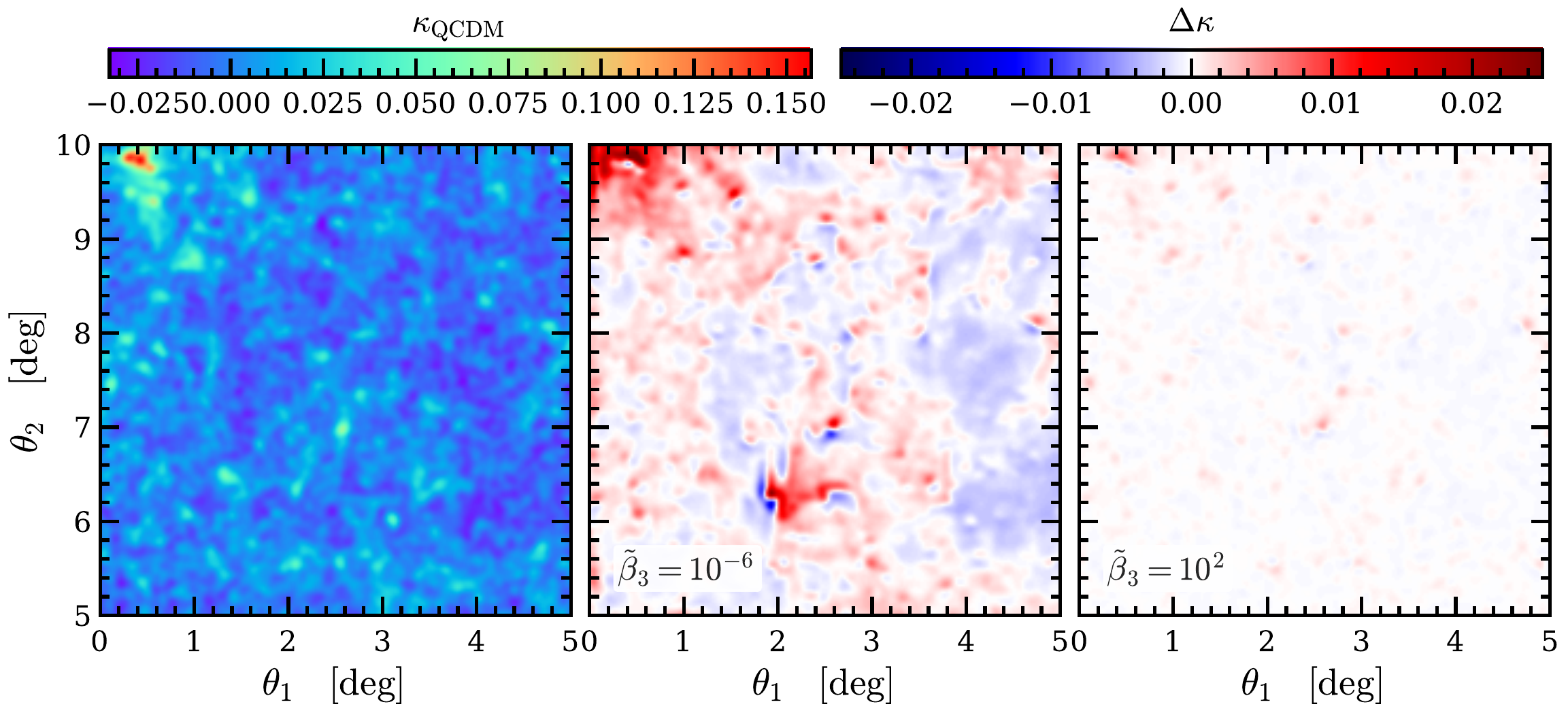}
    \caption{We visualise a portion of the \orig{} maps, smoothed by a Gaussian kernel with a width $\theta = 2.5 \, {\rm arcmin}$, of in QCDM (left), and the difference between the cvG model and their QCDM counterpart with $\Delta\kappa = \kappa({\betathree}) - \kappa_{\rm QCDM}$, for $\betathree = [10^{-6}, 10^{2}]$ (centre and right respectively). The maps show the ray tracing results for the redshift range $z=[0.08, 1.0]$.
    }
    \label{fig:Fig_lensing_maps}
\end{figure}

It is important to note here, that we solve the integral of Eq.~\eqref{eq:kappa_cvg} between $z=[0.08, 1.0]$, as we found that artefacts appear for the $\betathree=10^{-6}$ cvG model. The reason behind this might be explained through the failure of numerical computation of the $\tilde{\chi}$ field in under-dense regions. This is a problem which has been reported multiple times \cite{Barreira:2013eea,Li:2013tda,Barreira:2013xea,Winther:2015pta} and discussed in terms of the cvG model in \cite{Becker:2020azq}.

\begin{figure}[!b]
    \centering
    \includegraphics[width=.98\textwidth]{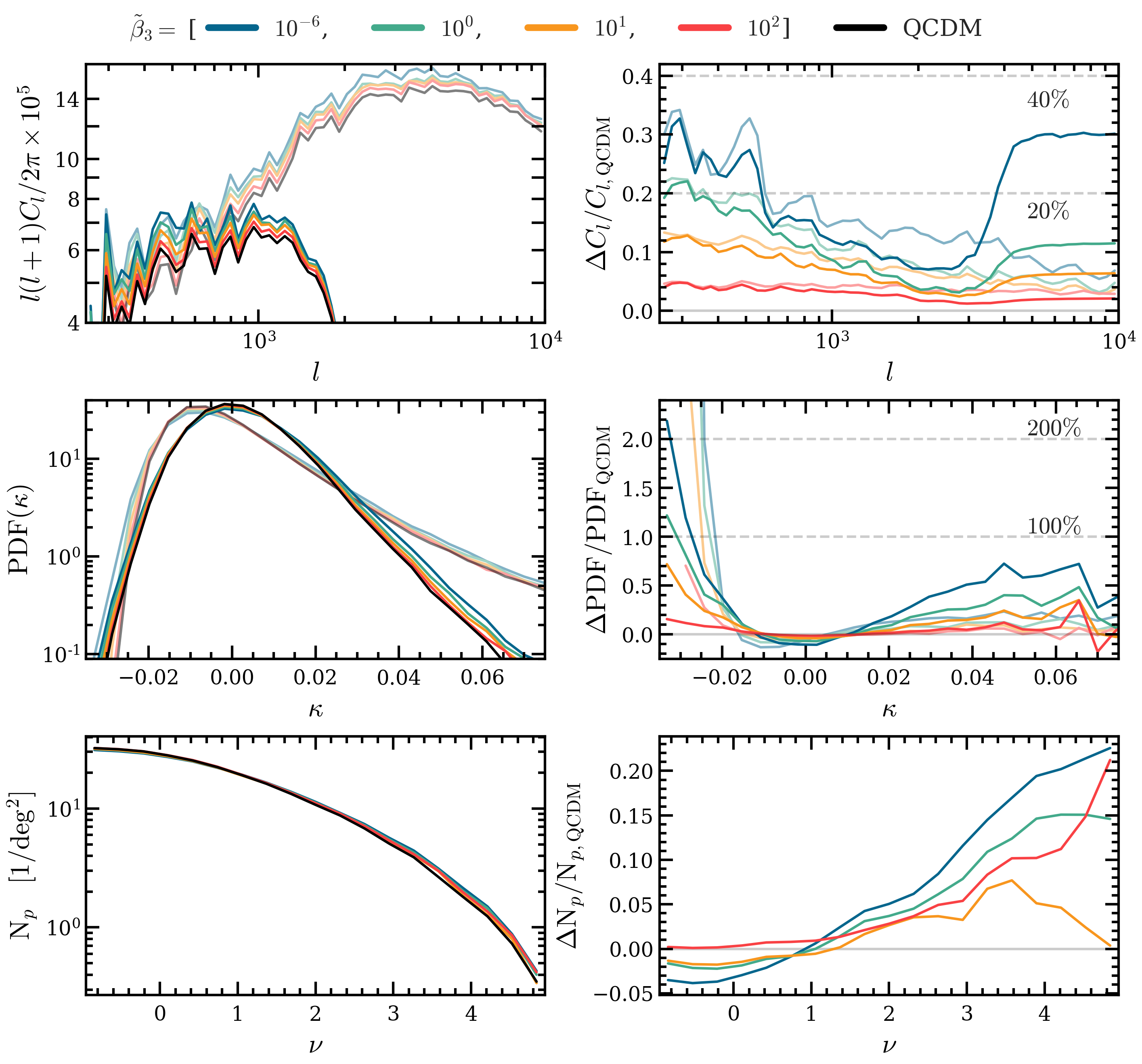}
    \caption{Weak lensing statistics: lensing convergence angular power spectra ({\it top}), probability distribution function of the weak lensing convergence field ({\it middle}), weak lensing peak abundance plotted as a function of peak height ({\it bottom}). The results shown here are obtained using a $10 \times 10$ deg$^2$ partial sky-map for a redshift range $z=[0.08, 1.0]$. We show results of the \orig{} maps (faint) and the \orig{} maps including the galaxy shape noise map, \ngsn{}, and smoothed with a Gaussian kernel of width $\theta = 2.5 \, {\rm arcmin}$ (bright) for the cvG model variants (colour) and their QCDM (black) counterpart.
    }
    \label{fig:Fig_lensing}
\end{figure}

The resulting \orig{} map is shown in Fig.~\ref{fig:Fig_lensing_maps} for QCDM (left), together with the residual between QCDM and the cvG model, $\Delta\kappa = \kappa({\betathree}) - \kappa_{\rm QCDM}$, for $\betathree = [10^{-6}, 10^{2}]$ (centre and right respectively). All maps have been smoothed with a Gaussian kernel of width $\theta = 2.5 \, {\rm arcmin}$ which we will abbreviate as \sg. It is clearly visible how underdense and overdense regions are more pronounced for $\betathree \to 0$ while for $\betathree \to \infty$ the model approaches the behaviour of the QCDM cosmology.

In the middle panel of Fig.~\ref{fig:Fig_lensing_maps} we can see a number of `dipole' features, where a positive-residual `hot spot' ($\Delta\kappa>0$) is aligned with a `cold spot' ($\Delta\kappa<0$). This is produced by the transverse (i.e., perpendicular to the line of sight) motion of the halo which contributes most for a given line of sight: for this case the $\kappa$ peak in the left panel would have moved slightly, causing this dipole feature in the residual map. Such dipoles are harder to find in the right panel, again because for $\betathree\rightarrow0$ the model behaves very similarly to QCDM, so that haloes move little compared with the latter case.

Another feature worth mentioning in the middle panel of Fig.~\ref{fig:Fig_lensing_maps} is that we can see that near the massive structures the convergence field is enhanced by over $10\%$. This is partly due to the increased halo masses, but most likely the dominant effect here is the fact that the Proca field can also modify the lensing potential, as mentioned above. While we shall not investigate it here, let us note that this means that weak lensing by galaxy clusters can be a potential probe to constrain this model. However, as in the case of csG \cite{Barreira:2015fpa}, we expect that the constraining power of cluster lensing may be limited by Vainshtein screening in the vicinity of clusters. We shall see shortly that this strong enhancement of convergence can be detected in the convergence power spectrum (or the shear correlation function) which can probe large-scale variations of the lensing potential.

In observations, the WL signal is obtained by averaging the shearing of source galaxy shapes over a large number of source galaxies whose intrinsic ellipticity dominates over the physical tangential shear signal. This effect is known as galaxy shape noise (GSN) and is a main source of uncertainty on small angular scales. We include the GSN by modelling it as a Gaussian random field which we will denote as \ngsn{}. Therefore we assume that \ngsn{} is independent of the underlying \orig{}. Furthermore, we assume that the correlation function of \ngsn{} is a $\delta$ function, thus pixel values show no correlation. The standard deviation of the Gaussian distribution is given by
\begin{equation}\label{eq:gsn}
    \sigma^2_{\rm pix} = \frac{\sigma^2_{\rm int}}{2 \theta_{\rm pix} n_{\rm gal}},
\end{equation}
where $\sigma_{\rm int}$ is the intrinsic ellipticity dispersion of the source galaxies, $\theta_{\rm pix}$ is the width of each pixel, and $n_{\rm gal}$ is the measured source galaxy number density. We use $\sigma_{\rm int}=0.4$ and $n_{\rm gal} = 40 \, {\rm arcmin}^{-2}$, which match LSST specifications \cite{Abell:2009aa}.

In the top row of Fig.~\ref{fig:Fig_lensing} we show the results for the power spectrum of the \orig{} maps (faint) and the \orig-\ngsn-\sg{} maps (bright). We do not include the linear theory prediction, as it holds up to $\ell\lesssim10^2$ and is thus outside of the range of multipoles we are able to extract from the maps. The left panel shows the absolute power spectra measurements for which we have not included the results for $\ell > 10^4$ as such small angular scales are not well-resolved given our simulation resolution. In terms of the relative difference between the cvG models to their QCDM counterpart in the right panel, the curves show the expected behaviour that, on large angular scales $(\ell < 10^4)$, the amplitude is higher in the cvG models with smaller $\betathree$. However, since we use a partial-sky map of $10 \times 10 \deg^2$, the power spectra in the left panel could suffer from a large sample variance. This, however, should not strongly affect the result of the relative difference, as it roughly cancels out. As we go to smaller angular scales, $l\rightarrow10^4$, all cvG models converge toward their QCDM counterpart, which reflects the operation of the screening mechanism on small scales, e.g., inside haloes. Note that the smoothed maps behave similarly, though not identically, to the unsmoothed ones at $\ell\lesssim10^3$, while on smaller angular scales the smoothing significantly changes the model difference. This indicates a potential limitation on using the convergence power spectrum or shear two-point correlation function to test the cvG model, but we note that the large angular scales are where the model difference is most prominent anyway.

The middle row of Fig.~\ref{fig:Fig_lensing} shows the one-point distribution of the \orig{} maps (faint) and the \orig-\ngsn-\sg{} maps (bright). It contains information on non-Gaussian aspects of the convergence field that are not included in the convergence power spectra. We can see that cvG models with smaller $\betathree$ have larger numbers of pixels with both high and low $\kappa$ values. This behaviour is as expected because the fifth force in the cvG models helps to move more matter towards (from) dense (underdense) regions, as can be seen in Fig.~\ref{fig:Fig_lensing_maps}. It is good to see that increasing the $\betathree$ parameter indeed leads a smooth transition to QCDM, which is what is needed to cure the problem of having too strong a lensing effect in the csG model. The same happens to the void $\gamma_t$ profiles too, as will be shown in the next subsection.

The bottom row of Fig.~\ref{fig:Fig_lensing} shows the WL peak abundance for the \orig-\ngsn-\sg{} maps. This result is
useful on its own because WL peak statistics can be a useful cosmological probe (e.g.,  \cite{Jain:1999nu,Dietrich:2009,Shirasaki:2016twn,Li:2018owg,Liu:2016nfs,Davies:2019tjh,Fong:2019ixg,Coulton:2019enn}) but will also be useful for the study of void identified through WL peaks in the next subsection. We identify peaks as pixels whose $\kappa$ values are larger than those of their eight neighbours. For consistent definitions between the different cosmological models, we define the amplitude of $\nu$ of a map pixel as
\begin{equation}\label{eq:nu}
    \nu = \frac{\kappa}{\sigma_{\rm GSN}},
\end{equation}
where $\sigma_{\rm GSN}=0.007$ is the standard deviation of the \ngsn-\sg{} map generated using the LSST specifications given above. From the bottom panels of Fig.~\ref{fig:Fig_lensing}, we can see that for $\betathree \to 0$, there is a significant increase in the numbers of the high-amplitude peaks, which indicates that the fifth force strongly enhances the lensing signal of these pixels (note that the fifth force also increases the halo masses as found in Fig.~\ref{fig:halo_mass_fct}, which also contributes to this). On the other hand, the abundance of small peaks ($\nu<1$) is reduced as $\betathree\rightarrow0$, because some of the haloes that produce peaks with $\nu<1$ in QCDM have been able to produce peaks with $\nu>1$ in the cvG models. This trend agrees qualitatively with results found for the nDGP cosmology \cite{Davies:2019yif}.

\subsection{Cosmic voids}
\label{subsec:voids}

Cosmic voids are regions in the Universe where the densities of dark matter or tracers are low. In recent years it has been shown that voids (e.g.,  \citep{Platen:2011,Clampitt:2014gpa,Gruen:2015jhr}) can be a useful probe for a variety of models (e.g., \cite{Bos:2012wq,Cai:2014fma,Barreira:2015vra,Pisani:2015jha,Barreira:2016ias,Hamaus:2016wka,Higuchi:2016ucj,Falck:2017rvl,Cautun:2018ts,Paillas:2018wxs,Davies:2019yif,Baker:2018mnu,Pisani:2019cvo,Baker:2019gxo,Hamaus:2020cbu}), including the test of modified gravity models that are featured by Vainshtein screening \cite{Barreira:2015vra,Barreira:2016ias,Falck:2017rvl,Baker:2018mnu}. There are a large number of methods to find voids, and it has been argued that void identification based on WL convergence maps can lead to the better constraints of modified gravity theories \cite{Davies:2019yif}. This has motivated us to use voids from the two dimensional convergence field through the tunnel and watershed algorithms as the resulting void catalogues have been shown to be amongst the most promising \cite{Davies:2020xem}.

Whilst the convergence profiles of voids allow for a simpler physical interpretation of the mass content, where positive and negative $\kappa$ correspond to projected over-dense and under-dense regions, it is the tangential shear which can be measured directly in observations. Therefore, to offer a more straightforward comparison with observations, we study the void tangential shear profile $\gamma_t(r)$, which is related to the convergence profile through
\begin{equation}\label{eq:tangential_shear}
    \gamma_{t}(r) = \bar{\kappa}(< r) - \kappa(r),
\end{equation}
where
\begin{equation}\label{eq:kappa_bar}
    \bar{\kappa}(< r) = \frac{1}{\pi r^2} \int^{r}_{0} 2\pi r'\kappa(r') dr',
\end{equation}
is the mean enclosed convergence within radius $r$.

\subsubsection{Tunnels}
\begin{figure}[!b]
    \centering 
    \includegraphics[width=.98\textwidth]{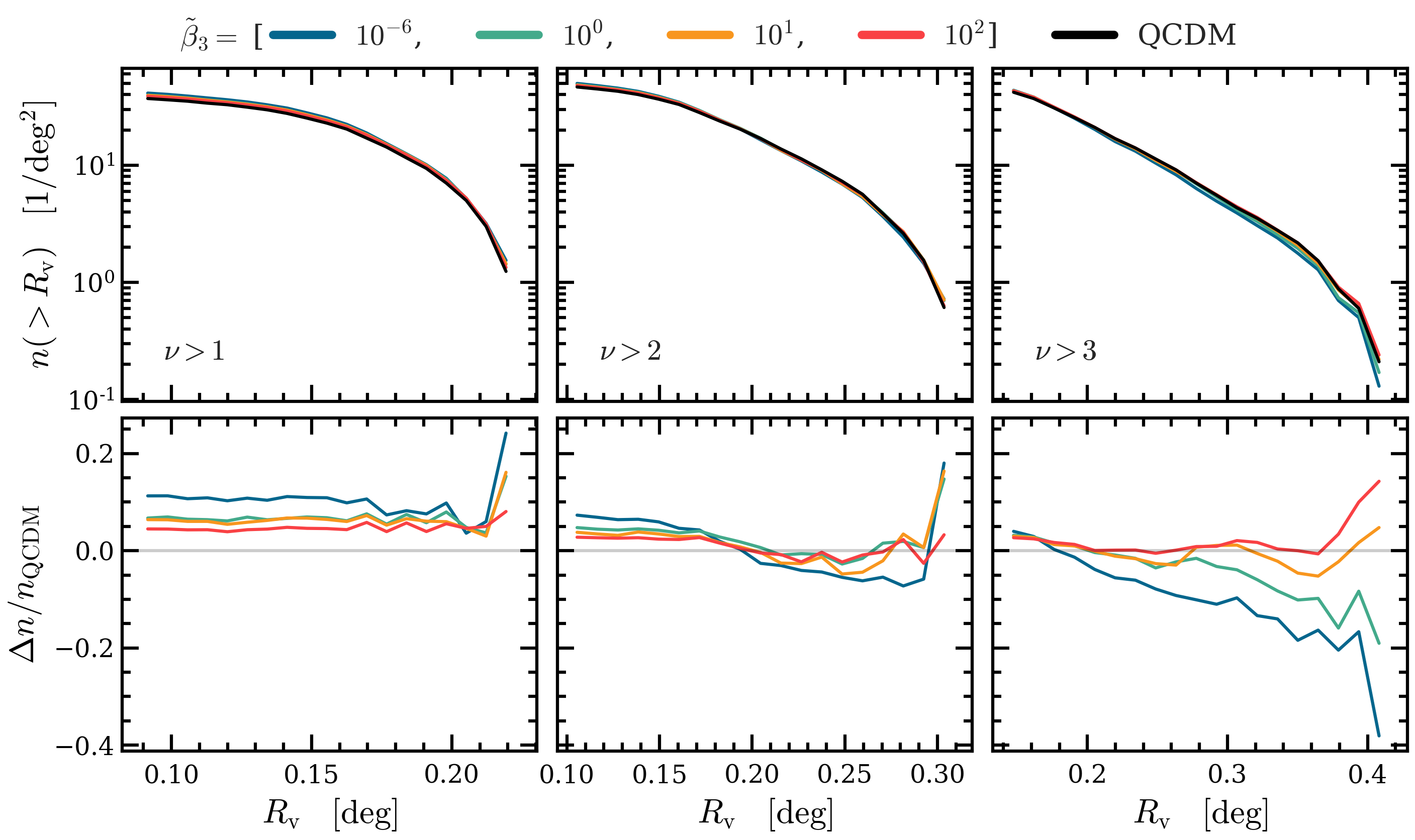}
    \caption{{\it Top}: the tunnel abundance as a function of their radii for the three WL peak categories: {\it left}: $\nu > 1$, {\it centre}: $\nu > 2$, {\it right}: $\nu > 3$. {\it Bottom}: relative difference between the cvG cosmologies and their QCDM counterpart.
    }
    \label{fig:Fig_void_size_fct}
\end{figure}

The tunnel algorithm  of \cite{Cautun:2018ts,Davies:2018jpm,Davies:2019yif} identifies voids based on a WL peaks catalogue. We will from now on refer to these voids as tunnels. We find peaks using the \orig\ map smoothed by a compensated Gaussian kernel wither an inner kernel width of $\theta_{\rm inner} = 2.5 \, {\rm arcmin}$ and a outer kernel width of $\theta_{\rm outer} = 15 \, {\rm arcmin}$, which we will abbreviate as \scg. The use of \scg{} instead of \sg{} is motivated by the larger number of identified peaks, which again will results in more identified tunnels and thus better statistics. Each identified peak is placed into three categories based on Eq.~\eqref{eq:nu}: $\nu > [1, 2, 3]$. For each category, a Delaunay tessellation with the peaks at the vertices is constructed. This produces a tessellation of Delaunay triangles, with a peak at the corner of each triangle, and no peaks within the triangles. Each Delaunay triangle is then used to construct its corresponding circumcircle, with the three vertices of the triangle falling on the circumcircle's circumference. This unique tessellation, by definition, produces circles which do not enclose any peaks. In order to increase the number of tunnels, which is necessary because of the small area of our convergence maps, we use all possible tunnels, including neighbouring ones which have a large degree of overlap in our study.

\begin{figure}[!b]
    \centering 
    \includegraphics[width=.98\textwidth]{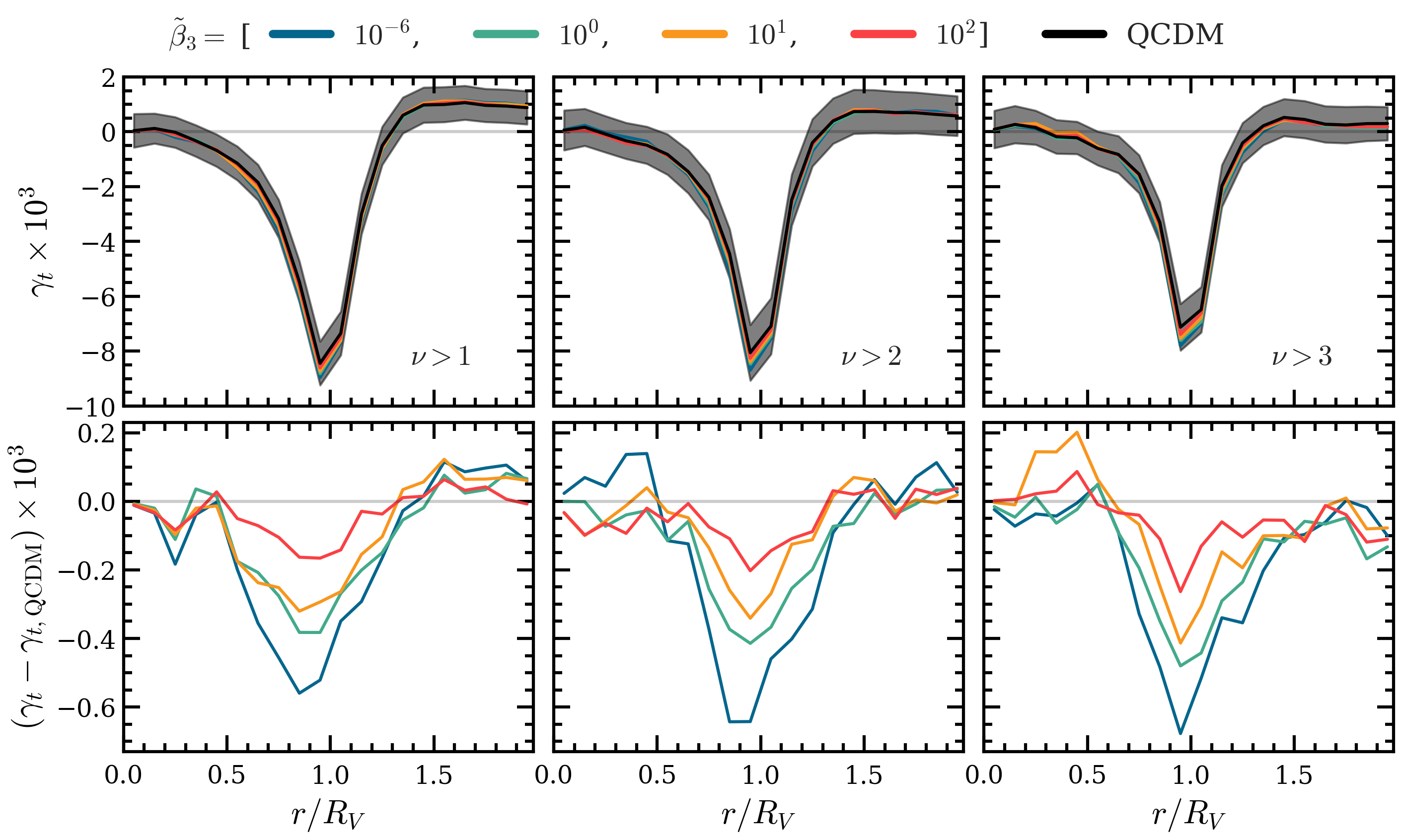}
    \caption{{\it Top}: Tunnel tangential shear profiles as a function of the scaled distance from the centre, $r/R_{\rm v}$, for QCDM (black) and cvG models with $\betathree=10^{-6}$ (blue), $10^0$ (green), $10^1$ (orange) and $10^2$ (red). The shaded region indicates the standard deviation of all tunnels in the QCDM map (for clarity we do not show this for the other models). The shaded region indicates the standard deviation. {\it Bottom}: The relative difference between the cvG models and their QCDM counterpart. From left to right the panels are respectively for tunnels identified from peak catalogues with peak height $\nu>1$, $2$ and $3$. We do not show the standard deviation as they very large due to our small sample size.
    }
    \label{fig:Fig_voids_tunnel_shear_profile}
\end{figure}

The top row of Fig.~\ref{fig:Fig_void_size_fct} shows the tunnel size distribution identified from peak catalogues of different significance: $\nu > 1$ (left), $\nu > 2$ (centre), and $\nu > 3$ (right). The smallest tunnels are generated by the $\nu > 1$ peak catalogue, which also produces the most tunnels, because the large number of peaks in this catalogue tends to partition the map into smaller Delaunay triangles. As the $\nu$ threshold increases, the typical tunnel size increases, however there are also fewer tunnels overall. This implies that each of the three categories should respond differently to the large scales modes of the \orig{} map, and thus creating the tightest constraints through combined analyses. Due to our small sample size, this remains to be tested.

The bottom row of Fig.~\ref{fig:Fig_void_size_fct} shows the relative difference between the cvG models and their QCDM counterpart. It is interesting to observe, that while smaller tunnels ($R_{\rm v} \lesssim 0.2 \, \deg$) are more abundant in cvG with $\betathree \to 0$ than in QCDM it is vice versa for larger voids ($R_{\rm v} \gtrsim 0.2 \, \deg$). This is a consequence of a higher abundance in WL peaks for the cvG cosmologies compared to their QCDM counterpart for all of our peak categories, see Fig.~\ref{fig:Fig_lensing}, more small voids and fewer large voids are found in cvG than in QCDM.

Fig.~\ref{fig:Fig_voids_tunnel_shear_profile} shows the tangential shear profiles, Eq.~\eqref{eq:tangential_shear}, of the three tunnel catalogues shown in Fig.~\ref{fig:Fig_void_size_fct}. The profile are based on the \orig-\ngsn{} maps, as smoothing would dampen the void profiles and the differences between the cosmological models. We compute the $\gamma_t$ profiles statistics by stacking all voids in a given catalogue, weighting them depending on their size (the smaller the void, the less its statistical weight). To obtain the 1-$\sigma$ error, indicated by the shaded region in the top row, we loop through 100 bootstrap resamples. We recover the typical tangential shear profile, which indicates that voids act as concave lenses. The extrema of the profile is located at $r \approx R_{\rm v}$ for all void categories and is increasing as the void sizes increase.

In the bottom row of Fig.~\ref{fig:Fig_voids_tunnel_shear_profile} we can clearly see that the potential well get deeper as $\betathree \to 0$, reflecting the effects of enhanced structure formation and modified photon geodesics. We do not show the bootstrapped 1-$\sigma$ error for the relative differences, as our sample size is too small.

\subsubsection{Watershed}

\begin{figure}[!b]
    \centering 
    \includegraphics[width=.98\textwidth]{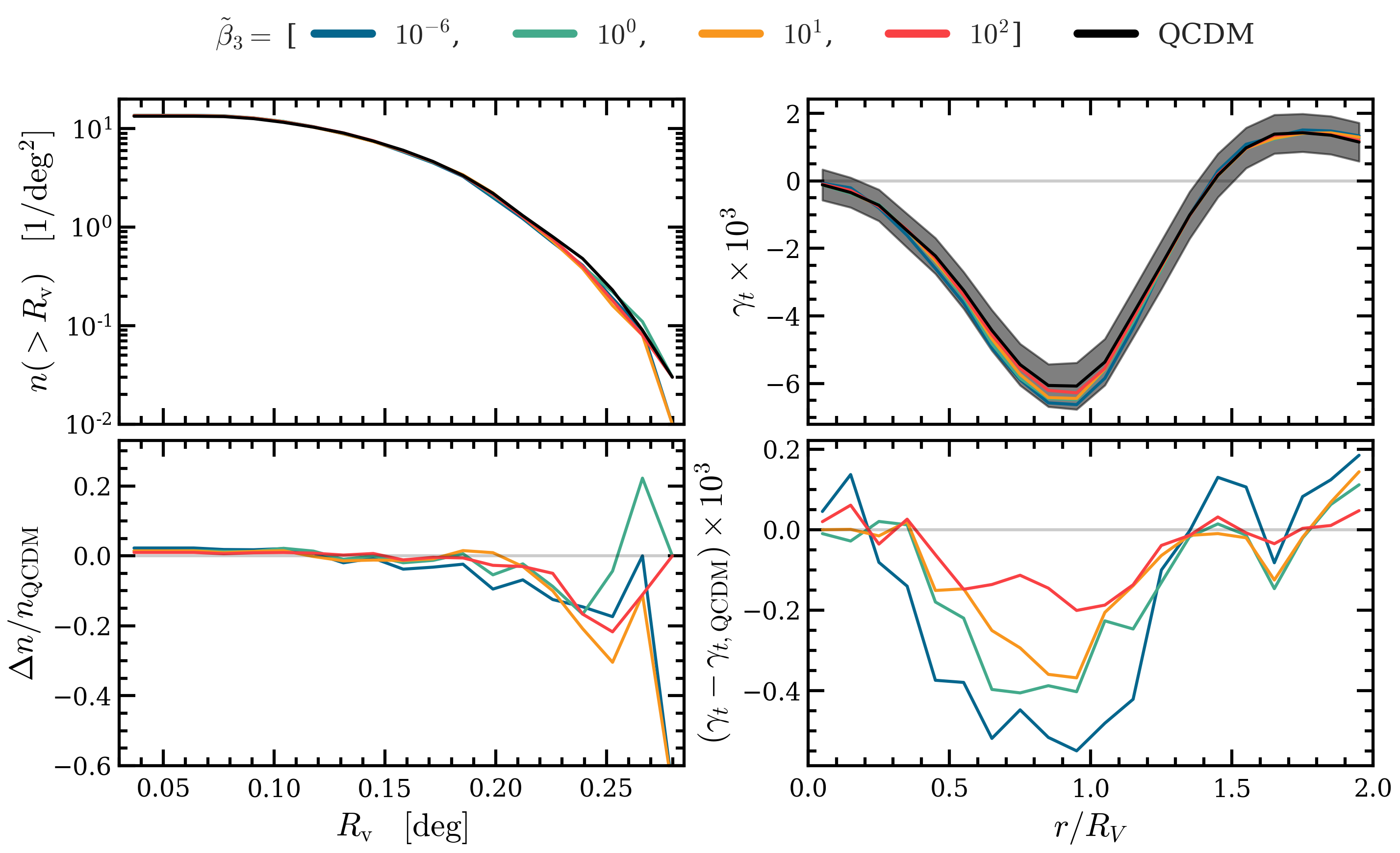}
    \caption{Statistics for the watershed voids. {\it Left}: the cumulative void abundance as a function of the effective  radius of the watershed voids, $R_{\rm v}$. {\it Right}: the tangential shear signal of these voids, as a function of the scaled radius from void centre, $r/R_{\rm v}$. The upper panels show the results for QCDM (black) and cvG models with $\betathree=10^{-6}$ (blue), $10^0$ (green), $10^1$ (orange) and $10^2$ (red), while the lower panels show the relative (for the void abundance) and absolute (for the tangential shear profile) differences between the cvG models and their QCDM counterpart. The shaded region in the top row indicates the standard deviation of all profiles in the QCDM model.
    }
    \label{fig:Fig_voids_watershed}
\end{figure}

The watershed algorithm of \cite{Platen:2007ar} identifies voids based on the basins in the topographic map which is constructed from the \orig{} map. To find the watershed basins, each pixel of the \orig{} map is connected to its neighbour with the lowest $\kappa$ value -- a process that is repeated for successive neighbours until a local minimum emerges. All pixels connected to the same minimum in this way form one watershed basin, with ridges of local high $\kappa$ values along the basin boundary. We could have used the WL peak catalogues to identify watershed voids, as is done for tunnels, but the results are generally very noisy \cite{Davies:2020xem}. To mitigate the impact of GSN, \cite{Davies:2020xem} found that the basin boundary should have a minimum \orig{} value of $\sigma_{\rm GSN} / 2$, as it allows watershed basins that have been artificially split by spurious structures introduced by GSN to be re-merged. Unlike tunnels, the watershed voids are formed by a collection of Delaunay cells, and therefore have irregular shapes. We define the void centre to be the barycentre of all selected cells for a given watershed void, and the void radius $R_{\rm v}$ as the radius of a sphere whose volume is equal to that of the void. The watershed algorithm has the advantage of simplicity from fewer free parameters in the void identification process, since no tracers are used, multiple WL peak catalogues do not need to be defined. However, Ref.~\cite{Davies:2020xem} also find that tangential shear profile from the watershed algorithm is more susceptible to GSN than the tunnel algorithm.

The left column of Fig.~\ref{fig:Fig_voids_watershed} shows the watershed void abundance as a function of the void radius, $R_{\rm v}$, and the relative difference between the cvG models and QCDM. In contrast to tunnels, there are overall fewer watershed voids, and they never reach the large void size as tunnels do. This is because watershed voids by definition cannot overlap. Among the different models, little difference is found, apart from the large-$R_{\rm v}$ end, where the cvG models produce up to $\sim20\%$ fewer voids than QCDM. The main reason for this is a change of void sizes, rather than a decrease in their number. This is likely due to the enhanced $\kappa$ field magnitude in local overdensities residing in larger underdense regions, which means that these structures would more easily have $\kappa>\sigma_{\rm GSN}/2$ and therefore become basin boundaries in the cvG models, leading to a split of a large waterbasin into smaller ones.

The right column of Fig.~\ref{fig:Fig_voids_watershed} shows the tangential shear profiles, $\gamma_t(r)$, of watershed voids and their relative difference between the cvG models and their QCDM counterpart. They are smoother, wider, and shallower compared to all tunnel categories. However, both tunnels and watershed voids reach their tangential shear profile minimum at $0.9-1.1\, R_{\rm v}$. The error bars on the QCDM tangential shear profiles from the two algorithms are also similar in size, which suggests that both algorithms may offer similar constraining power, consistent with Ref.~\cite{Davies:2020xem} which finds roughly similar tangential shear signal-to-noise ratios between the two algorithms. The relative differences between the cvG models and their QCDM counterpart peak at the minimum of the profile, with a $10\%$ difference for cvG with $\betathree = 10^{-6}$, roughly the same as the relative difference found for tunnels in the same size range (which is the tunnel category for $\nu > 1$).

\section{Discussion and conclusions}
\label{sec:conclusions}

In this paper, we have performed a thorough phenomenological study of a simplified version of the generalized Proca theory, the vector Galileon model (cvG). To study the impact of the cvG models free parameter, $\betathree$, we have run a set of five realizations of simulations for $\betathree = [10^{-6}, 10^{0}, 10^{1}, 10^{2}]$ and their QCDM counterpart, resulting in a total of $25$ simulations. The study relied on an adapted version of the \ecosmog{} N-body code augmented with the ray-tracing modules of the \rayramses{} algorithm. We used the five independent realisations for each model to create a light cone that covers a field of view of $10 \times 10 \deg^2$ from $z=0.08$ to a source redshift of $z=1$ (cf.~Sec.~\ref{sec:simulations_section} and Fig.~\ref{fig:lc_setup}). This allows us to study the matter, halo and weak lensing statistics. In the following we shall summarise the results of each those three topics.

The study of dark matter field statistics finds good agreement with \cite{Becker:2020azq} about the {\it matter power spectrum} ($P_{\delta\delta}$, cf.~Sec.~\ref{subsec:power_spectra} and Fig.~\ref{fig:Pk_matter}), but extends the results of that paper by including larger scales and showing statistical uncertainties. In addition:

$\bullet$ the simulation measurements of the {\it velocity divergence power spectrum} ($P_{\theta\theta}$, cf.~Sec.~\ref{subsec:power_spectra} and Fig.~\ref{fig:Pk_div_velo}) converge to the linear-theory prediction on scales $k \lesssim 0.1 \hMpc$ for all times, while for $k \gtrsim 0.1 \hMpc$ we reproduce the well-known result that $P_{\theta\theta}$ is suppressed compared to the linear theory results.
The relative difference, $\Delta P_{\theta\theta}(k)/P_{\theta\theta,{\rm QCDM}}(k)$, shows that the wavenumber at which linear theory and simulation results agree reasonably, $k_\ast$, is pushed to ever larger scales as $a \to 1$ and $\betathree \to 0$. Finally, for $a \to 1$ and $\betathree \to 0$ we see a growing peak that for the case of $\betathree = 10^{-6}$ protrudes above the linear theory prediction at $k \sim 0.7 \, \hMpc$. A similar feature was also observed by \cite{Li:2013nua} for the DGP model.

$\bullet$ for the {\it matter bispectrum} ($B$, cf.~Sec.~\ref{subsec:bispectra} and Figs.~\ref{fig:Fig_bispectrum_B}, \ref{fig:Fig_bispectrum_Q_R}), we find that the magnitudes depend on the triangle configurations, and increase in the order of equilateral, squeezed, and folded triangle configurations. However, this order is reversed when considering the relative difference. The relative difference confirms that, as it is the case for $P_{\delta\delta}$ and $P_{\theta\theta}$, the tree-level bispectrum is a good estimator on large scales $k<k_\ast \sim 0.1 \, \hMpc$, while the exact value of $k_\ast$ decreases with $a \to 1$ and $\betathree \to 0$. We show that the enhancement of the bispectrum due to the fifth force is marginally stronger than in the case of power spectrum, but the reduced bispectrum shows that $B/B_{\rm QCDM}$ is to a very good approximation equal to $\left(P/P_{\rm QCDM}\right)^2$. The scales at which we are able to measure the bispectrum do not show a strong signature of the Vainshtein screening.

The study of halo statistics is mostly based on \subfind{} cagalogues, as they contain the smallest haloes and subhaloes and thus can enable measurements to smaller scales, although where possible we have also cross-validated the results with \fof{} haloes. The main observations are the following:

$\bullet$ the {\it halo mass function} ($n(>M)$, cf.~Sec.~\ref{subsec:halo_mass_functions} and Fig.~\ref{fig:halo_mass_fct}) shows that the fifth force enhances the abundance of dark matter haloes in the entire mass range probed by the simulations, with the enhancement stronger at late times and for high-mass haloes. Models with a weaker fifth force, e.g., with $\betathree\rightarrow\infty$, show a more restrained enhancement of the HMF.

$\bullet$ the {\it two-point correlation function} ($\xi(r)$, cf.~Sec.~\ref{subsec:correlation_functions} and Fig.~\ref{fig:Fig_two_point_corr_fct}) shows more strongly enhanced clustering for smaller values of $\betathree$, for which the fifth force is stronger. The enhancement of the halo $\xi(r)$ is nearly constant down to $\sim 3 \, \Mpch$, consistent with $P_{\delta\delta}$, and reflecting the fact that in the cvG model the growth factor is enhanced in a scale-independent way in the linear regime. However, the enhancement in halo clustering is weaker than in matter clustering, for all models at all times.

$\bullet$ the relative difference of the {\it mean halo pairwise velocity} ($v_{\langle ij \rangle}$, cf.~Sec.~\ref{subsec:mean_pairwise_velocity} and Fig.~\ref{fig:Fig_pairwise_velocity_00}) remains constant for all cvG models at scales $r > 10 \, \Mpch$, in very good agreement with linear-theory prediction. For the latter, we have measured the halos bias, $b$, for four different scale factors through the relation between the halo and matter correlation functions. The resulting measurements of $b$ for the different models are similar, but show a slight decrease as $\betathree \to 0$, as the fifth force enhances matter clustering more than halo clustering, as mentioned above.

$\bullet$ the {\it redshift space halo clustering} ($\xi_{\ell}(s)$, cf.~Sec.~\ref{subsec:redshift_space_clustering} and Fig.~\ref{fig:Fig_tpcf_s}) is sensitive to the halo pairwise velocity and hence the fifth force. The relative difference between cvG and QCDM can be up to $\sim 3$ times larger for the quadrupole, $\xi_2(s)$, than for the monopole, $\xi_0(s)$, although its SNR is $\sim 0.5$ times smaller on the range $20 < s \, \hMpc < 40$ due to larger statistical uncertainly in the halo velocity field. Future data of redshift space distortions should provide strong constraints on $\betathree$.

$\bullet$ the result of the {\it halo concentration-mass relation} ($c_{200}$,  cf.~Sec.~\ref{subsec:mass_concentration_relation} and Fig.~\ref{fig:concentration_mass_rel}) shows that in the cvG model, just as for the csG model, the Vainshtein screening is strong enough inside haloes that there is little effect of the fifth force.

Our final section concerns the properties of the weak lensing convergence, peak and void statistics, where voids are identified using the tunnel and watershed algorithms. The main results are the following:

$\bullet$ the difference of the {\it convergence map} (\orig{}, cf.~Sec.~\ref{subsec:weak_lensing} and Fig.~\ref{fig:Fig_lensing_maps}) between QCDM and cvG for $\betathree = 10^{-6}$ shows that around massive structures the convergence field is enhanced by over $10\%$. However, we caution about taking this as an indication that weak lensing by galaxy clusters can be a potential probe to constrain this model, as we have not performed an analysis of stacked weak lensing convergence profiles.

$\bullet$ the relative difference of the {\it angular power spectrum} ($C_{\ell}$, cf.~Sec.~\ref{subsec:weak_lensing} and Fig.~\ref{fig:Fig_lensing_maps}) is largest on linear scales $\ell \lesssim 3\times 10^2$, reaching $\sim30\%$ for $\betathree\rightarrow0$. These scales are also where the smoothing of the map has little impact on the relative difference. For higher multipoles the model differences reduce.

$\bullet$ the relative difference of the {\it probability distribution function of \orig{}} (${\rm PDF}(\kappa)$, cf.~Sec.~\ref{subsec:weak_lensing} and Fig.~\ref{fig:Fig_lensing_maps}) shows that cvG models with $\betathree \to 0$ have more pronounced under- and overdense regions.

$\bullet$ the relative difference of the {\it weak lensing peak abundance} ($N_p$, cf.~Sec.~\ref{subsec:weak_lensing} and Fig.~\ref{fig:Fig_lensing_maps}) shows larger (smaller) numbers of high- (low-)amplitude peaks for $\nu > 1$ ($\nu < 1$) in the cvG models with $\betathree\rightarrow0$, because the fifth force enhances the convergence values of the peak pixels.

$\bullet$ the relative difference of the tunnel and watershed {\it void abundances} ($N(>R_{\rm v})$, cf.~Sec.~\ref{subsec:voids} and Fig.~\ref{fig:Fig_void_size_fct}, \ref{fig:Fig_voids_watershed}) shows fewer large-sized voids in the cvG cosmologies compared to their QCDM counterpart, since they produce more weak lensing peaks which splits large voids into smaller ones (for the tunnel case), or increase the convergence values so that the regions satisfying the chosen void definition criterion shrink in size (for the watershed case).

$\bullet$ the relative difference of the {\it tangential shear profile} for tunnels and watershed voids (cf.~Sec.~\ref{subsec:voids} and Fig.~\ref{fig:Fig_voids_tunnel_shear_profile}, \ref{fig:Fig_voids_watershed}) peak at approximately the void radius, with up to $10\%$ difference for the cvG model with $\betathree = 10^{-6}$ (similar to what has been observed in the convergence maps), and the model difference decreases as $\betathree\rightarrow\infty$.

Overall, we find that for the cvG model studied here, the fifth force effect is strongest on velocity and lensing statistics. The former is because velocity is the first integration of acceleration, and thus reacts quickly to the enhancement of gravity due to the fifth force, which happens only at late times; the matter density field, in contrast, reacts more slowly as the second integration of acceleration. The latter is because in the cvG model, unlike for some other MG models, photon geodesics are affected in two different ways: (1) indirectly, by the modified growth of matter fluctuations, and (2) directly, by the fifth force. This suggests that redshift space distortions and weak lensing shear correlation functions can both be promising cosmological probes to constrain the $\betathree$ parameter in this model. On small scales, the models are generally more difficult to constrain because the screening mechanism suppresses the fifth force effect; for example, internal properties of haloes, such as the concentration-mass relation, are insensitive to the fifth force. Another potentially useful way to constrain this model is by cross-correlating galaxies with the integrated Sachs-Wolfe effect \cite{Nakamura:2018oyy}, because as $\betathree\rightarrow0$ the fifth force becomes stronger, causing the lensing potential to getting deeper rather than shallower \cite{Becker:2020azq} as suggested by observations. This possibility will be investigated in future. What is a bit surprising is that weak lensing by voids do not seem to be as promising a probe, even though the lensing potential is significantly modified in low-density regions: perhaps this is because weak lensing is a cumulative effect along the line of sight, and this strong effect in low-density regions is somehow cancelled out by the weaker effects in high-density regions.

Recently, various studies to constrain the generalized GP theory using cosmological observations have been conducted, see, e.g., \cite{deFelice:2017paw,DeFelice:2020sdq,Heisenberg:2020xak}. These studies focused on general nonlinear functional forms for $G_{2,3}$, because linear forms of these functions, such as the models studied here, have been found as a poor fit to observational data. However, as suggested by \cite{Becker:2020azq}, adding massive neutrinos with significantly nonzero mass (see, e.g., \cite{Barreira:2014ija}) may be a way to make the GP model with linear $G_{2,3}$ agree better with data. This possibility will be studied in a follow-up work, and correspondingly we hope to include massive neutrinos in future simulations.

\acknowledgments

CB and CTD acknowledge support by the UK Science and Technology Facilities Council (STFC) PhD studentship through a Centre for Doctoral Training and grant ST/R504725/1 respectively. AE and BL are supported by the European Research Council (ERC) through Starting Grant ERC-StG-716532-PUNCA. BL additionally acknowledges support by the STFC through grants No.~ST/T000244/1 and ST/P000541/1. This work used the DiRAC@Durham facility managed by the Institute for Computational Cosmology (ICC), on behalf of the STFC DiRAC HPC Facility (www.dirac.ac.uk). This equipment was funded by BEIS capital funding via STFC capital grants ST/K00042X/1, ST/P002293/1, ST/R002371/1 and ST/S002502/1, Durham University, and STFC operations grant ST/R000832/1. DiRAC is part of the UK National e-Infrastructure.

\bibliographystyle{JHEP}
\bibliography{bibliography} 

\end{document}